\documentclass[
 preprint,12pt]{cls/elsarticle}

\usepackage{graphicx}
\usepackage{dcolumn}
\usepackage{bm}
\usepackage{amsmath}
\usepackage{amssymb}

\def\C{\mathbb{C}}
\def\R{\mathbb{R}}
\def\N{\mathbb{N}}
\def\Z{\mathbb{Z}}

\def\rP{\mathrm{P}}
\def\var{\mathrm{var}}
\def\ro{\mathrm{o}}
\def\rS{\mathrm{S}}
\def\rI{\mathrm{I}}
\def\rR{\mathrm{R}}

\def\bP{\mathbf{P}}

\def\cG{\mathcal{G}}
\def\cN{\mathcal{N}}

\def\bra{\langle}
\def\ket{\rangle}

\journal{} 

\begin{document}

\begin{frontmatter}



\title{Continuum Percolation and Stochastic Epidemic Models \\
on Poisson and Ginibre Point Processes}

\author[inst1]{Machiko Katori}

\affiliation[inst1]{%
organization={Department of Information Physics and Computing, 
Graduate School of Information Science and Technology, 
The University of Tokyo}, 
addlessline={Hongo, Bunkyo-ku, Tokyo 113-0033, Japan}}
\ead{katori-machiko@g.ecc.u-tokyo.ac.jp}

\author[inst2]{Makoto Katori}

\affiliation[inst2]{%
organization={Department of Physics, Faculty of Science and Engineering, 
Chuo University}, 
addlessline={Kasuga, Bunkyo-ku, Tokyo 112-8551, Japan}}
\ead{katori@phys.chuo-u.ac.jp}

\begin{abstract}
The most studied continuum percolation model
in two dimensions
is the Boolean model consisting of disks with the same radius
whose centers are randomly distributed on the 
Poisson point process (PPP).
We also consider the Boolean percolation model on the
Ginibre point process (GPP) which is a typical
repelling point process realizing hyperuniformity. 
We think that the PPP approximates a disordered configuration
of individuals, while the GPP does a configuration of citizens
adopting a strategy to keep social distancing in a city
in order to avoid contagion.
We consider the SIR models with contagious infection 
on supercritical percolation clusters
formed on the PPP and the GPP.
By numerical simulations, we studied dependence of
the percolation phenomena and the infection processes
on the PPP- and the GPP-underlying graphs. 
We show that in a subcritical regime of 
infection rate the PPP-based models
show emergence of infection clusters 
on clumping of points which 
is formed by fluctuation of 
uncorrelated Poissonian statistics. 
On the other hand, 
the cumulative numbers of infected individuals
in processes are suppressed 
in the GPP-based models.
\end{abstract}



\begin{keyword}
continuum percolation model \sep
SIR model \sep
Poisson point process \sep
Ginibre point process \sep
social distancing \sep
infection clusters
\end{keyword}

\end{frontmatter}

\section{Introduction}
\label{sec:introduction}
The study of stochastic and
statistical-mechanics modeling of epidemics
has a long history, and a variety of
types of model has been proposed and
extensively studied
\cite{KM27,Bai53,Bai57,Lig85,AM91,Lig99,DH00,CHBC09,Bar16}.
We choose a type of model or decide to invent
a new model depending on the purpose
of analysis of infection processes using mathematical models. 
Here we are interested in \textit{contagious disease} of human beings
which is spreading in a city, so spatial structures should be
included in a model.
In usual modeling, we choose a graph and consider
that each site represents an individual and
edges between neighboring sites
do interactions between individuals.
For example, we fix a finite but large-scale subdomain
$\Lambda$ of the square lattice $\Z^2$
with some boundary condition, 
and to each site $x \in \Lambda$ put a random variable
$\eta(x)$, which takes one of the three states, being
susceptible (S), infected (I), or recovered (R) \cite{KM27}.
We consider the global configuration of individuals
$\eta_t :=\{\eta_t(x)\}_{x \in \Lambda}$
and define a stochastic process 
$(\eta_t)_{t \geq 0}$ by specifying transition rules
of random variables in continuous time $t \geq 0$. 
The main topic of study on such a 
\textit{lattice SIR model} 
\cite{Gra83,dST10,TZ10,SMDHB20,SAAMF20,Zif21} 
is to clarify the dependence of the time-evolution 
of global configuration $\eta_t$ 
on spacial structures and relevant parameters
specifying the transition rules
 (e.g., infection rate and recovering rate).

In a real city, however, 
when a serious contagious disease is spreading, 
citizens try to change their behavior in order to avoid contagion.
A typical strategy is to keep
\textit{social distancing}.
It is obvious that if all individuals keep
the distances from their neighbors be
greater than the range of contagion,
then the disease will be extinct.
The problem is however, that real individuals
are not fixed at sites of a regular lattice,
and hence they occasionally break social distancing
and form groups, and it causes a risk to make
\textit{infection clusters}. 

A system of points in a space which are
randomly distributed following a specified probability law
is generally called a \textit{point process}.
Note that this mathematical terminology
means a purely static system without any
time evolution \cite{DVJ03}.
The purpose of the present study is 
to introduce stochastic epidemic models
of individuals whose locations are 
given by a random point process, and
to clarify dependence of the infection processes
on the \textit{underlying graph} 
$\cG$ given by a point process.
\begin{figure}[ht]
\includegraphics[width=1\linewidth]{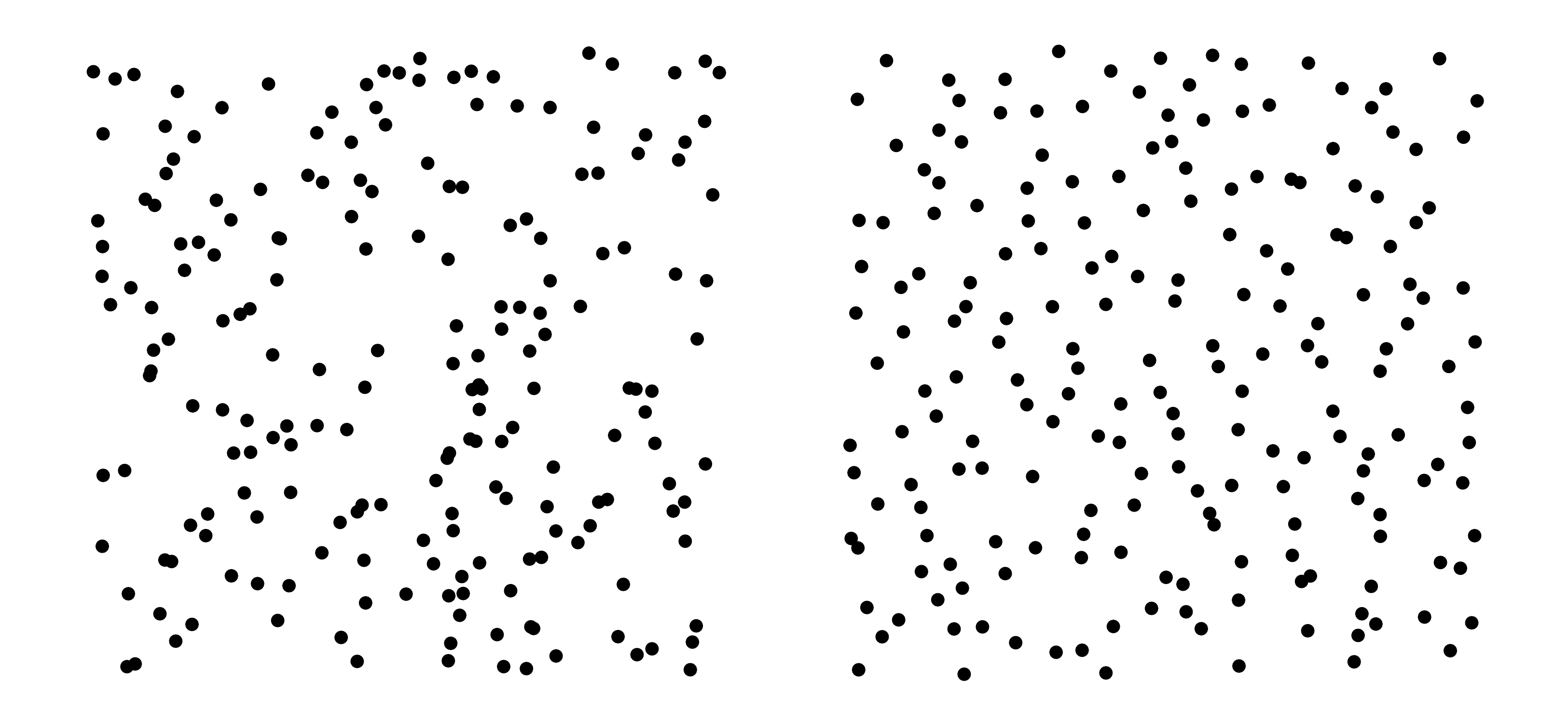}
\caption{%
Typical realizations of point configurations
in the Poisson point process (PPP) in the left,
and in the Ginibre point process (GPP) in the right.
In both figures, 
the total number of points is 200.
}
\label{fig:GPP_PPP}
\end{figure}

In the present paper we use two distinct
point processes on a plane $\R^2$
to define underlying graphs; 
 the Poisson point process (PPP) and
the Ginibre point process (GPP) \cite{Gin65}.
The left figure of Fig.~\ref{fig:GPP_PPP} 
shows a typical realization of 200 points 
of the PPP, while the right one shows that of the GPP.
The GPP is realized as a bulk scaling limit of
eigenvalue distributions on a complex plane $\C$
of Gaussian complex random matrices,
which has been extensively studied 
in random matrix theory \cite{Meh04,For10}.
It is a typical two-dimensional example 
of a large family of repelling point processes called the 
\textit{determinantal (fermion) point processes}
\cite{Sos00,ST03a,ST03b,Shi06,HKPV09,Kat15}. 
On the other hand, there is no correlation among
points in the PPP. In this sense, the PPP can be
said to be a uniform distribution of point process. 
We observe in the left figure of Fig.~\ref{fig:GPP_PPP}, 
however, that clumping of points and vacant spaces
happen to occur. 
In the PPP the variance of the number of points 
included in the disk 
with center $x \in \R^2$ and radius $r>0$, 
$B_r(x) : = \{y \in \R^2 : |y-x| < r\}$, 
is proportional to $r^2$; 
that is, the fluctuation
of the number of points in a disk is in the
same order with the total number of points in it.
In the GPP, due to the repulsive interactions 
among points, the variance of the number of points in 
$B_r(x)$ is only proportional to $r$ in $r \to \infty$. 
(Note that both point processes are translationally invariant
and then the center $x$ of disk is arbitrary
in the above assertions.) 
Such a suppression of number fluctuations 
is a common feature of deterministic lattices, 
determinantal point processes, and
other correlated particle systems, and called
\textit{hyperuniformity} \cite{Tor18,MKS21}.
The GPP will mimic locations of citizens in a city
who try to keep social distancing, 
while the PPP will do a society in which
such a strategy to avoid contagion is not adopted at all.

\begin{figure}[h]
\includegraphics[width=1\linewidth]{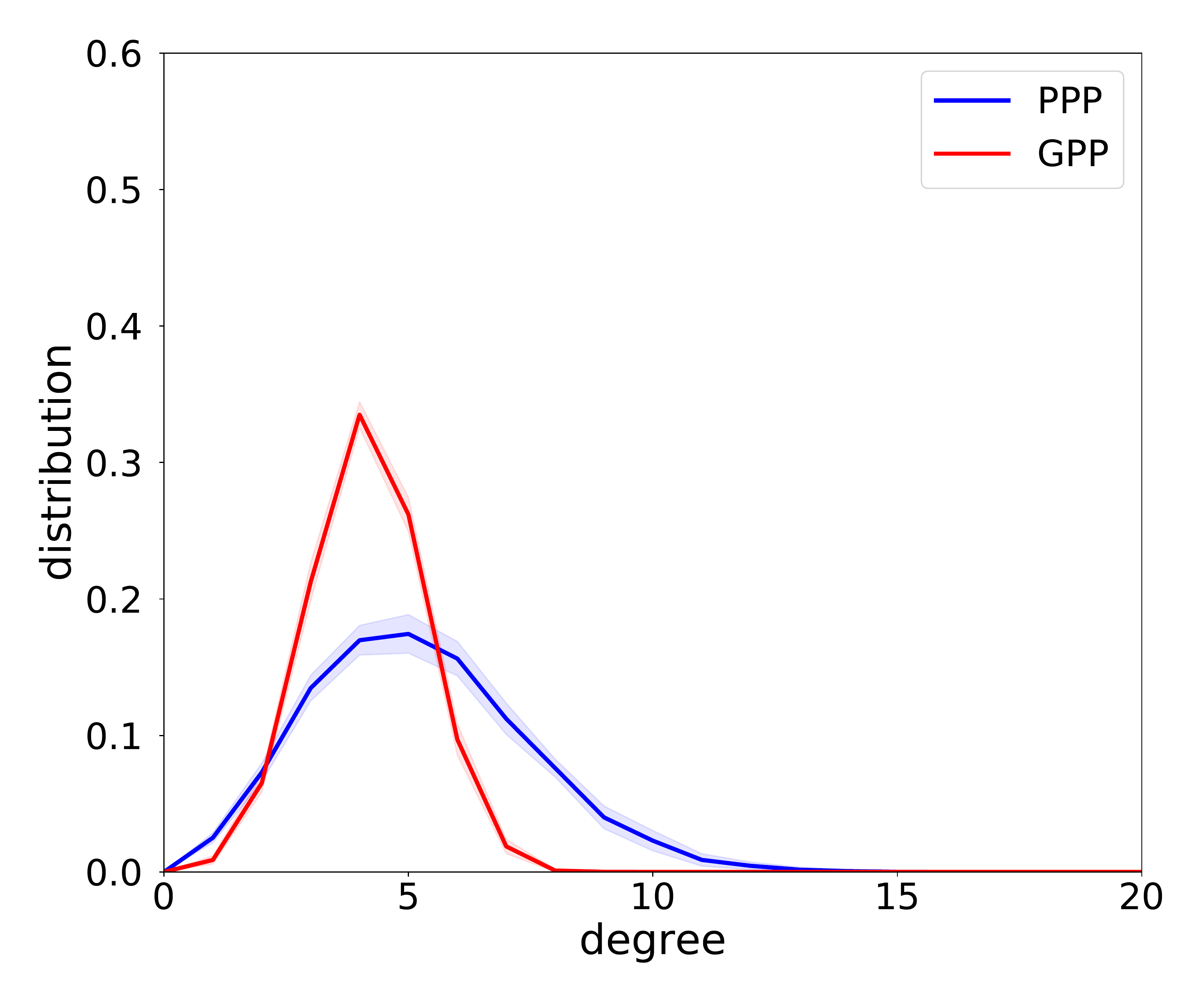}
\caption{
Distributions of degree of each points
in the PPP-underlying graph $\cG^{\rm PPP}$ and
the GPP-underlying graph $\cG^{\rm GPP}$.
Solid lines show the means of
ten distinct samples of random graphs and
shaded strips show the standard deviations of them.
The mean values of degree are given by
5.23 for $\cG^{\rm PPP}$ and
4.15 for $\cG^{\rm GPP}$. 
}
\label{fig:degree_distribution}
\end{figure}
We express the range of infection by 
a parameter $r >0$ 
and consider an infection process such that
only if the distance between two individuals
is less than $2r$, contagion can occur.
In other words, the maximum set of individuals
who have the possibility to be infected is given by
a set of points $x$ each of which has at least
one neighboring point $y$ such that
$B_r(x)\,{\cap}\,B_r(y)\not=\emptyset$.
Such a set is known as a percolation cluster
of the \textit{Boolean percolation model}
or the
\textit{standard Gilbert disk model} \cite{Gilb61}
in the continuum percolation theory
\cite{MR96,BR06,BY13,BY14,GKP16}.
We have numerically generated typical samples
of large \textit{percolation clusters} 
on the PPP and the GPP and
regarded them as a pair of distinct 
underlying graphs
$\cG^{\rm PPP}$ and $\cG^{\rm DPP}$
for our epidemic models.
In order to compare essential differences between them,
we have imposed the condition 
on the Boolean percolation models 
from which $\cG^{\rm PPP}$ and 
$\cG^{\rm GPP}$ are obtained such
that the filling factors, 
defined by (\ref{eqn:filling_factor}) below, 
take the same value $\kappa$.
The specified value of $\kappa$ is
greater than the critical values of
percolation transitions 
on the PPP and on the GPP;
$\kappa > \kappa_{\rm c}^{\rm PPP} 
> \kappa_{\rm c}^{\rm GPP}$. 
One of the differences between these two
graphs is represented by the distributions
of \textit{degrees} of each point,
where the degree of each point means
the number of neighboring points in the graph.
As shown by Fig.~\ref{fig:degree_distribution},
points with larger values of degrees are realized in 
$\cG^{\rm PPP}$ compared to $\cG^{\rm GPP}$.

We have numerically studied the SIR models \cite{KM27}
on $\cG^{\rm PPP}$ and on $\cG^{\rm GPP}$
starting from the same initial configuration with only one
infected (I) individual at a randomly chosen point
in a field of susceptible (S) individuals.
The infection process $\rS \to \rI$ takes place
with rate $\lambda \Psi(n)$ and the
recovering $\rI \to \rR$ with rate 
$\mu \equiv 1$, where the infectivity
$\lambda >0$, $\Psi$ is a positive function,
and $n$ gives the total number of neighbors
who are infected. The recovered (R) individuals
are stable and never to be infected again.
If $\lambda$ is large, infection spreads
more efficiently on $\cG^{\rm GPP}$ than on $\cG^{\rm PPP}$
due to the hyperuniformity of the GPP.
In such a case, the strategy such as
keeping social distancing does not work. 
\begin{figure}[h!]
\includegraphics[width=1\linewidth]{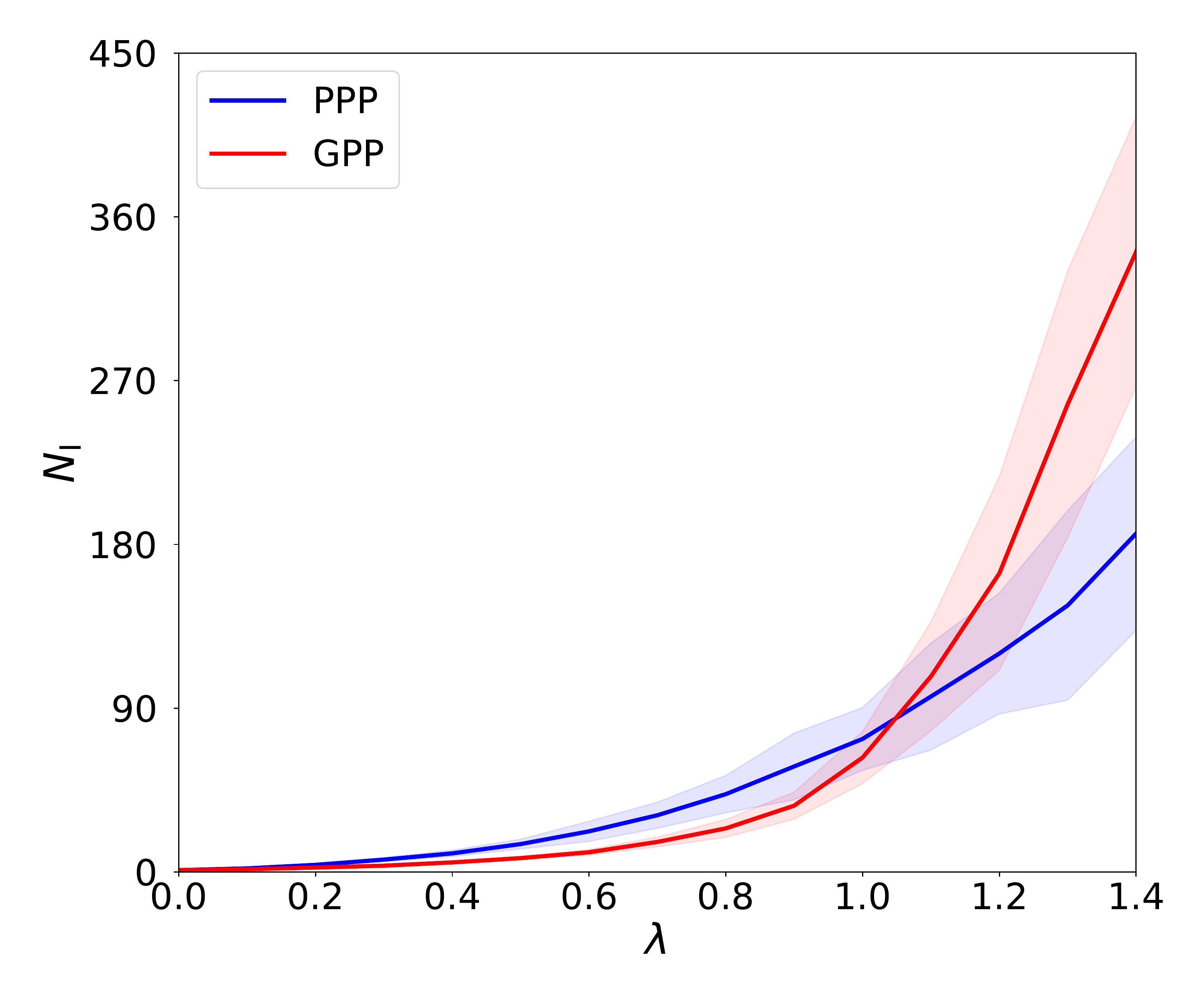}
\caption{%
For each underlying graph $\cG$ we performed
100 runs of the SIR models in which
the infection rate is given by
$\lambda \widehat{\Psi}(n)$ with a quadratic function
$\widehat{\Psi}(n)=n^2$, 
and $n$ denotes the number of infected neighbors. 
We defined the cumulative number 
of infected individuals $\cN_{\rI}$ 
by the mean value of the 100 runs.
The quenched averages $\bra \cN_{\rI} \ket$
over ten distinct samples of $\cG$ are
shown versus $\lambda$ for
the PPP-based model by a blue curve
and for the GPP-based model by a red curve.
The standard deviations in the ten samples are shown by
shaded strips around the curves.
We find a crossing point at $\lambda_* \simeq 1.0$
of these two curves.
}
\label{fig:I_vs_lambda_quadratic}
\end{figure}

\begin{figure}[h]
\begin{minipage}{0.45\hsize}
\includegraphics[width=1.2\textwidth]{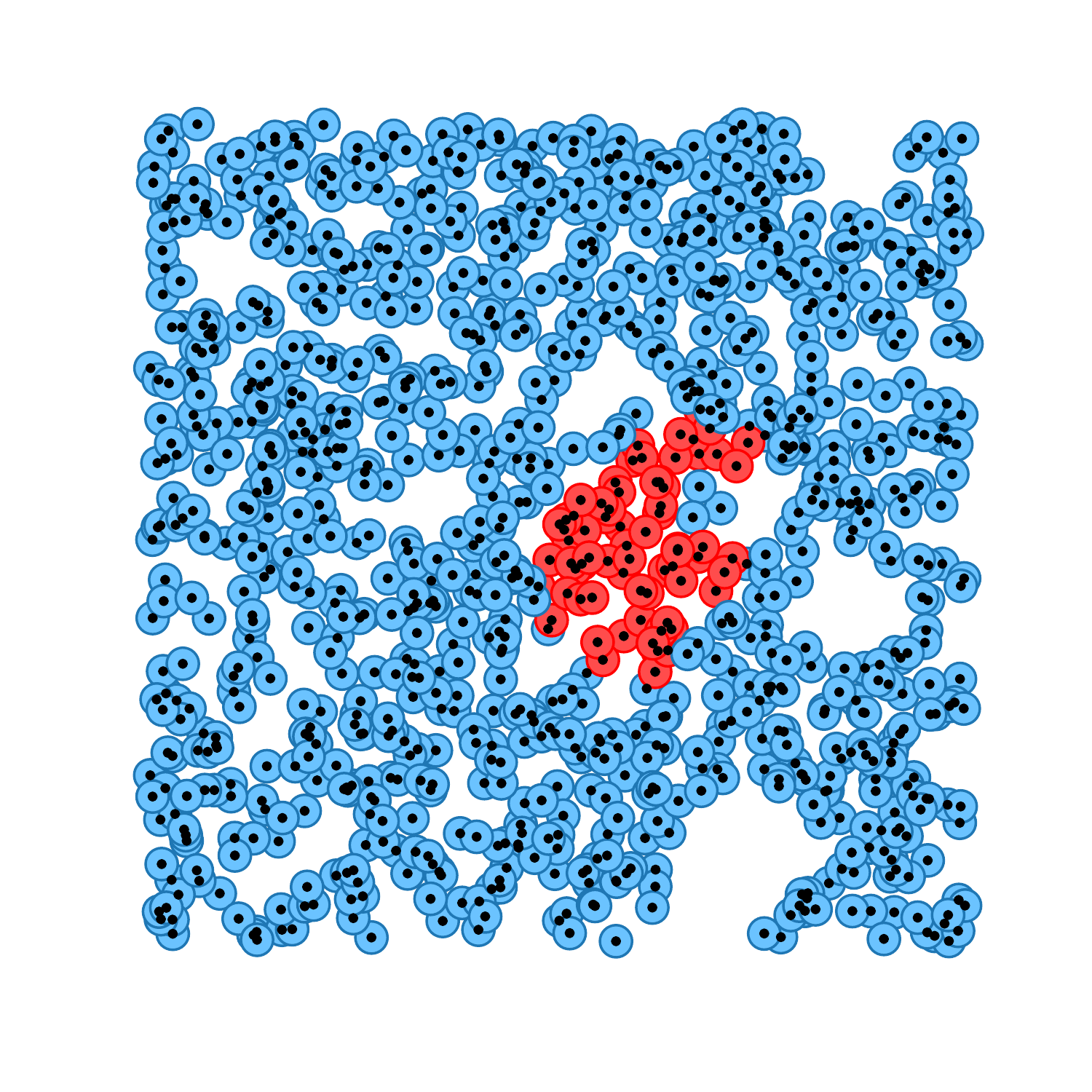}
\end{minipage}
\begin{minipage}{0.45\hsize}
\includegraphics[width=1.2\textwidth]{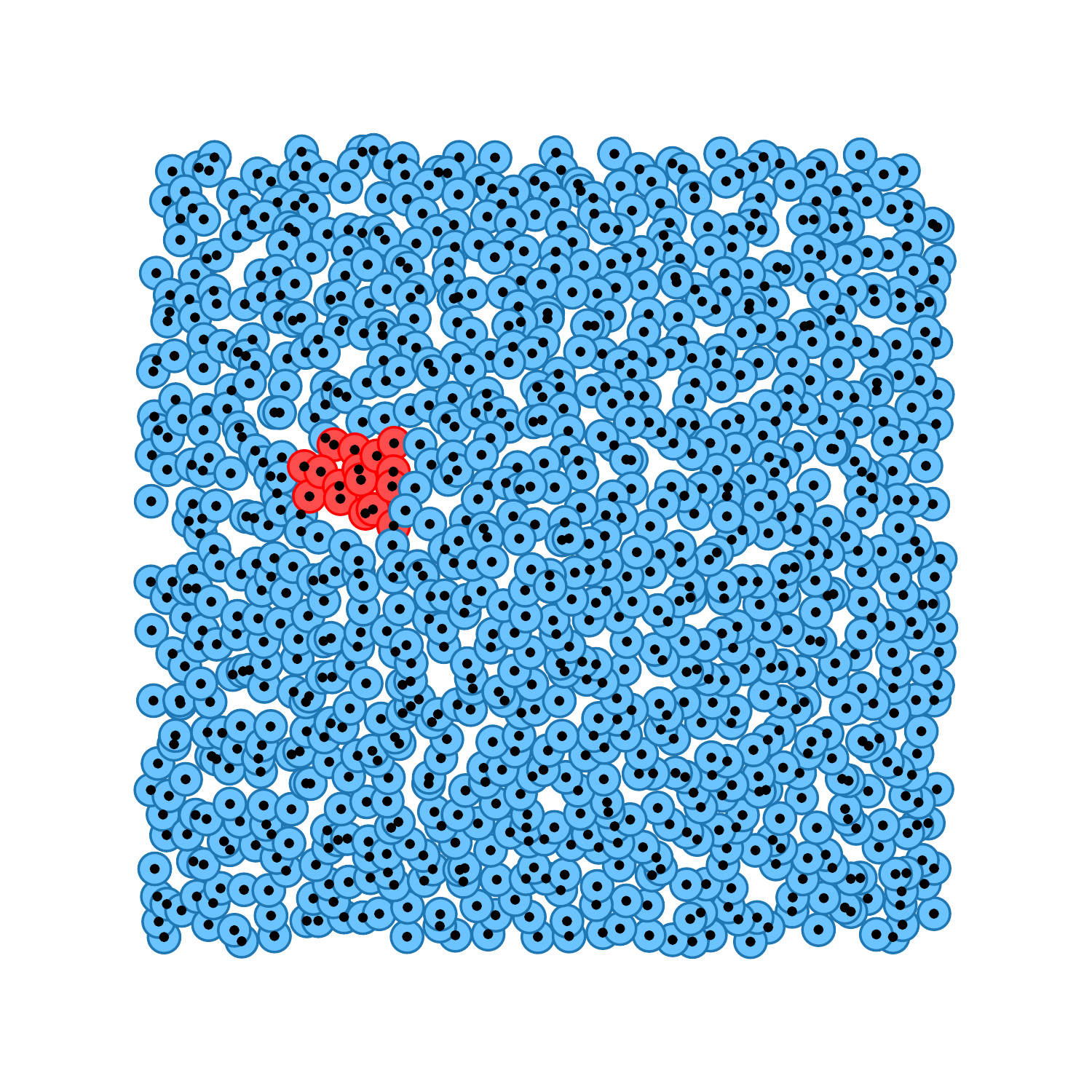}
\end{minipage}
\caption{%
Examples of final configuration of the SIR model
on $\cG^{\rm PPP}$ in the left
and on $\cG^{\rm GPP}$ in the right.
Here $\lambda=0.8$ and the quadratic function
$\widehat{\Psi}(n)=n^2, n \in \N$ is assumed. 
Blue and red disks show 
$\rS$- and $\rR$-individuals, respectively.
The numbers of $\rR$-individuals
in the final configurations,
which are equal to $\cN_{\rI}$, 
are 68 in the left and 18 in the right, respectively.
}
\label{fig:infection_clusters}
\end{figure}

In the present work, we are interested in the situation
in which $\lambda$ is relatively small and
\textit{infection clusters} do not spread over
large-scale. 
The main assertion of the preset paper is that
in this case the cumulative number of infected individuals
$\cN_{\rI}$ in an infection process can be
suppressed in the GPP-based model
compared to the PPP-based model.
It was confirmed that this phenomenon is realized
in the simple `linear model' in which
$\Psi(n)=n, n \in \N:=\{1,2,\dots\}$, and then
we studied its dependence on the functional form
of $\Psi(n)$.
Figure \ref{fig:I_vs_lambda_quadratic} shows 
dependence of $\cN_{\rI}$ on $\lambda$
for the `quadratic model' in which we assume 
$\widehat{\Psi}(n)=n^2, n \in \N$.
In each sample of $\cG$, the SIR model
with a given value of $\lambda$ was simulated 
100 times using the Gillespie algorithm \cite{Gil76,Gil77,EC20}
and $\cN_{\rI}$ is defined by the mean value of 100 runs.
Then we evaluate the average 
$\bra \cN_{\rI} \ket$ and the standard deviation
over ten distinct samples of $\cG$.
This procedure should be regarded
as \textit{quenched averaging} of
the SIR model on \textit{random environments}
given by underlying graphs $\cG$.
We find a special value $\lambda_* \simeq 1.0$
such that
$\bra \cN_{\rI}^{\rm PPP} \ket > \bra \cN_{\rI}^{\rm GPP} \ket$
if $0 < \lambda < \lambda_*$,
while 
$\bra \cN_{\rI}^{\rm PPP} \ket < \bra \cN_{\rI}^{\rm GPP} \ket$
if $\lambda > \lambda_*$.
Since $\lambda_*$ is much less than the critical
infection rate $\lambda_{\rm c}$, 
infection processes
cease sooner or later and final configurations
are expressed by a confined and 
connected domain consisting of
$\rR$-individuals embedded in 
a field of $\rS$-individuals.
Typical final configurations are
shown in Fig.~\ref{fig:infection_clusters}
for the PPP-based model in the left
and for the GPP-based model in the right.
We observe an accumulation of 
$\rR$-individuals on a clumping of points
in $\cG^{\rm PPP}$.
Emergence of such infection clusters
is suppressed in the model on $\cG^{\rm GPP}$
when $\lambda < \lambda_*$.
Remember that there is no correlation at all
in the PPP and clumping of points
is formed accidentally.
This implies that infection clusters can 
be caused by fluctuation of point processes
even though the infectivity $\lambda$ is relatively small.

The paper is organized as follows.
In Section \ref{sec:models} first we give the mathematical
definitions of the PPP and the GPP,
and then the Boolean percolation models on them are 
introduced.
We briefly explain the percolation transition
at the critical filling factor $\kappa_{\rm c}$, 
and we use an infinite percolation cluster which can 
appear in the supercritical phase $\kappa > \kappa_{\rm c}$
as a underlying graph $\cG$ for our
epidemic models.
We define the SIR models with
contagious infections on $\cG$.
We report our numerical results on 
the Boolean percolation models on
the PPP and the GPP 
in Section \ref{sec:results_percolation}.
Numerical estimations of
$\kappa_{\rm c}^{\rm PPP}$ and
$\kappa_{\rm c}^{\rm GPP}$ are given there.
Section \ref{sec:results_SIR} is devoted to
the simulation results of the SIR models
and dependence of 
infection processes on the two distinct
underlying graphs
$\cG^{\rm PPP}$ and $\cG^{\rm GPP}$ is studied.
We performed simulations of the SIR models
by modifying the function $\Psi$.
Dependence of the ratio
$\bra \cN_{\rI}^{\rm GPP} \ket/
\bra \cN_{\rI}^{\rm PPP} \ket$
on $\lambda$ is shown for each version of model,
which will help us to understand
the suppression mechanism of infection clusters 
in the GPP-based model
in the parameter regime $0 < \lambda < \lambda_*$.
Concluding remarks are listed out
in Section \ref{sec:concluding_remarks}.

\section{The Models and their Basic Properties}
\label{sec:models}
\subsection{The Poisson point process (PPP)}
\label{sec:PPP}

For a domain $A \subset \R^2$, 
its area is denoted by $|A|$.
So if $A$ is bounded, then $|A| < \infty$.
We write the Poisson point process (PPP) on $\R^2$
as $\Xi^{\rm PPP}$.
$\Xi^{\rm PPP}$ is a counting measure of 
randomly distributed points on $\R^2$ 
such that, for any bounded $A \subset \R^2$, 
the number of points included in $A$ 
is given by a random variable $\Xi^{\rm PPP}(A)$.
$\Xi^{\rm PPP}$ with intensity $\alpha >0$
satisfies the following \cite{DVJ03,MR96}.
\begin{enumerate}
\item
For mutually disjoint bounded domains $A_1, \dots, A_{m}$,
$m \in \N$;
that is, $A_j \cap A_k = \emptyset$, $\forall j \not= k$,
the random variables 
$\Xi^{\rm PPP}(A_1), \dots, \Xi^{\rm PPP}(A_{m})$
are mutually independent.

\item
For any bounded domain $A \subset \R^2$, 
\begin{equation}
\rP(\Xi^{\rm PPP}(A)=\ell)
=e^{-\alpha|A|} \frac{\alpha^{\ell} |A|^{\ell}}{\ell !},
\label{eqn:PPP1}
\end{equation}
$\ell \in \N_0:=\{0,1, \dots\}$.
\end{enumerate}

By (\ref{eqn:PPP1}), the mean value of $\Xi^{\rm PPP}(A)$
is calculated as
\begin{align*}
\bra \Xi^{\rm PPP}(A) \ket
&:= \sum_{\ell=0}^{\infty} \ell \rP(\Xi^{\rm PPP}(A)=\ell)
\nonumber\\
&=e^{-\alpha |A|} \sum_{\ell=0}^{\infty}
\ell \frac{\alpha^{\ell} |A|^{\ell}}{\ell !}
= \alpha |A|, 
\end{align*}
for $A \subset \R^2$ with $|A| < \infty$,
and hence the density $\rho^{\rm PPP}$ is equal to
the intensity,
\[
\rho^{\rm PPP} :=
\frac{\bra \Xi^{\rm PPP}(A) \ket}{|A|} = \alpha.
\]
Similarly, the variance of the number of points
included in a bounded domain $A \subset \R^2$ is calculated as
\[
\var(\Xi^{\rm PPP}(A)) :=
\bra(\Xi^{\rm PPP}(A)-\bra \Xi^{\rm PPP}(A) \ket)^2 \ket 
=\alpha |A|,
\]
which implies that as $|A| \to \infty$
the variance grows
in the same order of the mean of the
number of points,
\[
\lim_{|A| \to \infty}
\frac{\var(\Xi^{\rm PPP}(A))}{\bra \Xi^{\rm PPP}(A) \ket} =1.
\]
 
\subsection{The Ginibre point process (GPP)}
\label{sec:GPP}

Here we identify the real two-dimensional
plane $\R^2$ with a complex plane $\C$.
For $z = x+ i y \in \C$ with
$i :=\sqrt{-1}$, $x, y \in \R$, we write
$\overline{z} := x-i y$, 
$dz = dx dy$.
For $z=x+iy, z'=x'+iy' \in \C$,
$z \overline{z'}= (xx'+yy')-i(xy'-x'y)$;
in particular $|z|^2:= z \overline{z}=x^2+y^2$.

Consider an $L \times L$ matrix $M_L=(m_{jk})_{1 \leq j, k \leq L}$,
$L \in \N$, whose entries are
independently and identically distributed (i.i.d.)
complex Gaussian random variable with
mean zero and variance $\sigma^2$;
\begin{equation}
\rP(m_{jk} \in dz)
=\frac{1}{\pi \sigma^2} e^{-|z|^2/\sigma^2} dz,
\quad j, k =1,2, \dots, L.
\label{eqn:cNormal}
\end{equation}
We can prove that the eigenvalues of $M_L$,
$\{Z_j\}_{j=1}^L$ in uniform random order, obey 
the following joint probability distribution,
\begin{align*}
&\rP(Z_1 \in dz_1, \dots, Z_L \in dz_L)
\nonumber\\
& \quad 
=\frac{1}{\pi^L \sigma^{L(L-1)} \prod_{\ell=1}^L \ell !}
e^{-\sum_{\ell=1}^L |z_{\ell}|^2/\sigma^2}
\nonumber\\
& \qquad \times
\prod_{1 \leq j < k \leq L}
|z_k-z_j|^2 dz_1 \cdots dz_L, 
\end{align*}
$z_1, \dots, z_L \in \C$.
The factor $\prod_{1 \leq j < k \leq L} |z_k-z_j|^2$ 
gives repulsive correlation between any pair of points.

Such a correlated point process is generally characterized
by \textit{correlation functions} \cite{Meh04,For10,Kat15}.
For a point process $\Xi$, 
the $n$-th correlation function $\rho_n$,
$n \in \N$, 
is defined as the following.
Again we consider mutually disjoint bounded domains
$A_1, \dots, A_m \subset \R^2$, $m \in \N$.
For any set of integers
$k_1, \dots, k_m \in \N_0$
satisfying $\sum_{\ell=1}^m k_{\ell}=n \in \N_0$, 
\begin{align*}
&\left\bra
\prod_{\ell=1}^m \frac{\Xi(A_{\ell})}{(\Xi(A_{\ell})-k_{\ell})!} 
\right\ket
\nonumber\\
& \quad =\int_{A_1^{k_1} \times \cdots \times A_m^{k_m}}
\rho_n(z_1, \dots, z_n) dz_1 \cdots dz_n,
\end{align*}
where if $\Xi(A_{\ell})-k_{\ell} < 0$,
we interpret 
$\Xi(A_{\ell}) !/(\Xi(A_{\ell})-k_{\ell})!=0$. 
For any $L \in \N$, the eigenvalues $\{Z_j \}_{j=1}^L$ 
provides a determinantal point process (DPP) 
in the sense that any correlation function 
is expressed by a determinant
\[
\rho^L_n(z_1, \dots, z_L)
= \det_{1 \leq j, k \leq L}
[K_L(z_j, z_k) ],
\]
$n \in \{1,2, \dots, L\}$.
All correlation functions are 
specified by a two-point function
called the \textit{correlation kernel},
and in the present case it is given by \cite{Gin65}
\[
K_L(z, z')=\frac{1}{\pi \sigma^2}
e^{-(|z|^2+|z'|^2)/2 \sigma^2}
\sum_{\ell=0}^{L-1} \frac{1}{\ell !}
\left( \frac{z \overline{z'}}{\sigma^2} \right)^{\ell}.
\]

As $L \to \infty$, this function converges to 
\[
K^{\rm GPP}(z, z') = \frac{1}{\pi \sigma^2}
e^{-(|z|^2+|z'|^2)/2\sigma^2 + z \overline{z'}/\sigma^2},
\]
$z, z' \in \C$, 
which is hermitian; 
$\overline{K^{\rm GPP}(z, z')}=K^{\rm GPP}(z', z)$.
This limit correlation kernel defines a DPP 
with an infinite number of points on $\R^2$
through the correlation functions
$\rho^{\rm GPP}_n(z_1, \dots, z_n)
=\det_{1 \leq j, k \leq n}[K^{\rm GPP}(z_j, z_k)]$,
$n \in \N$. 
This infinite point process is called 
the Ginibre point process (GPP) and denoted by
$\Xi^{\rm GPP}$
\cite{Gin65,Meh04,Shi06,HKPV09,For10}.

The GPP is a uniform point process with density
\[
\rho^{\rm GPP} =\rho_1^{\rm GPP}
=K^{\rm GPP}(z,z)
=\frac{1}{\pi \sigma^2}.
\]
The two-point correlation function is
given by
\begin{align*}
\rho^{\rm GPP}_2(z, z')
&= 
 \det \left[
\begin{array}{ll}
K^{\rm GPP}(z, z) & K^{\rm GPP}(z, z') \cr
K^{\rm GPP}(z', z) & K^{\rm GPP}(z', z') 
\end{array}
\right]
\nonumber\\
&= K^{\rm GPP}(z,z) K^{\rm GPP}(z',z')
-|K^{\rm GPP}(z, z')|^2
\nonumber\\
&=(\rho^{\rm GPP})^2-|K^{\rm GPP}(z, z')|^2,
\end{align*}
$z, z' \in \C$.
The inequality 
$\rho^{\rm GPP}_2(z, z') - (\rho^{\rm GPP})^2 < 0$
holds and it 
implies negative correlation; that is,
$\Xi^{\rm GPP}$ is a repelling point process.

It was proved by \cite{Shi06} that
\[
\var(\Xi^{\rm GPP}(B_r) ) \sim \frac{r}{\sqrt{\pi} \sigma}
\quad \mbox{as $r \to \infty$}.
\]
This implies that the GPP is hyperuniform
in the sense \cite{Tor18},
\[
\frac{\var(\Xi^{\rm GPP}(B_r))}
{\bra \Xi^{\rm GPP}(B_r) \ket}
\propto r^{-1} \to 0
\quad \mbox{as $r \to \infty$},
\]
as mentioned in Section \ref{sec:introduction}. 
See also \cite{MKS21}.

\subsection{The Boolean percolation model
and critical filling factor}
\label{sec:boolean}

Let $\Xi$ be a point process on $\R^2$.
So far $\Xi$ has been considered as a measure
of random points on $\R^2$;
for instance, given a bounded domain $A \subset \R^2$,
$\Xi(A)$ counts the number of points included in $A$.
Here we regard a point process 
simply as a set of points.
Corresponding to $\Xi$, we define
\begin{equation}
\widehat{\Xi} :=
\{ x \in \R^2 : \Xi(\{x\}) \geq 1\}.
\label{eqn:hatXi}
\end{equation}
Notice that, by negative correlation, 
there is no multiple points
in the GPP;
$\Xi^{\rm GPP}(\{x\}) \in \{0,1\}$,
with probability one \cite{HKPV09}.
So the inequality `$\geq$' in (\ref{eqn:hatXi})
can be replaced by the equality `$=$'
for the GPP.

Now we introduce a real number $r>0$.
For two points $x, y \in \widehat{\Xi}$, 
if the Euclidean distance $|x - y| < 2r$,
then we say that $x$ and $y$ are
\textit{neighbors} of each other.
We place disks of radius $r$ around each point; 
$B_r(x),  x \in \widehat{\Xi}$. 
Then if and only if
$B_r(x) \cap B_r(y) \not= \emptyset$,
the two points $x$ and $y$ are neighbors.
Two points $x$ and $y$ are \textit{connected}
if there is a finite sequence of points 
$x_0, x_1, \dots, x_n \in \widehat{\Xi}$ such that 
$x_0=x, x_n = y$ and
$x_{j+1}$ is a neighbor of $x_j$, $j=0,1,\dots, n-1$.
The above defines the
\textit{Boolean percolation model} on the
point process $\widehat{\Xi}$ with radius $r$
\cite{Gilb61,MR96,BR06,BY13,BY14,GKP16}. 

The maximal connected components are called
\textit{percolation clusters}.
The size of a percolation cluster
is the number of points in $\widehat{\Xi}$ included in the
percolation cluster.
We say that the system percolates
if there is at least one 
\textit{infinite cluster}, whose size is infinity \cite{SA92,BR06}. 
The uniqueness of the infinite cluster with probability one
was proved for the Boolean percolation 
on the PPP and the GPP \cite{MR96,BR06,GKP16}. 
The probability that the system percolates is called
the \textit{percolation probability} 
and denoted by $\Theta$. 
In the present Boolean percolation model on $\R^2$,
$\Theta$ is a function of the
product of the density $\rho$ of the 
point process $\widehat{\Xi}$ and the area of a disk
$a=\pi r^2$,
\begin{equation}
\kappa := \rho a = \rho \pi r^2,
\label{eqn:filling_factor}
\end{equation}
which is called the \textit{filling factor}
\cite{MM12}.
There is a unique \textit{critical value of
filling factor} $\kappa_{\rm c}$ such that
$\Theta(\kappa)=0$, 
if $\kappa \leq \kappa_{\rm c}$. 
In the supercritical phase
$\kappa > \kappa_{\rm c}$, 
$\Theta(\kappa)>0$, and 
in the vicinity of $\kappa_{\rm c}$, 
$\Theta(\kappa)$ behaves as \cite{SA92},
\begin{equation}
\Theta(\kappa) \simeq \mbox{const.} \times
(\kappa-\kappa_{\rm c})^{\beta}, 
\quad \kappa \gtrsim \kappa_{\rm c}
\label{eqn:beta1}
\end{equation}
with a \textit{critical exponent} $\beta$. 
It has been argued that continuum percolation
and lattice percolation 
will belong to the same universality class
\cite{GS81,BBA83}.
The critical exponent of the percolation models on
usual planar lattices
is known as $\beta=5/36=0.1388\cdots$ \cite{SA92}.

\subsection{The SIR models on an infinite percolation cluster}
\label{sec:SIR}

Now we construct epidemic models on percolation clusters.
By definition of neighbors given above, 
a sequence of contagion should be limited within 
the percolation cluster.
We set $\kappa > \kappa_{\rm c}$
and consider an infinite cluster $\cG$ 
as a underlying graph on which our epidemic model is constructed. 
As will be explained in Section \ref{sec:results_percolation}
below, we will perform numerical simulations on finite point
processes put on a finite two-dimensional region
with the periodic boundary condition.
In such a case we will use the largest cluster
found in the finite system
as an underlying graph.
If we consider the SIR models starting
from one infected individual 
with weak infectivity, $\lambda \ll \lambda_{\rm c}$,  
an infection region is confined
in a relatively small area.
Then the finite-size effect of underlying graph
will be irrelevant.

Assume that an underlying graph is given by $\cG$. 
At each point $x \in \cG$, we put a random
variable $\eta(x) \in \{\rS, \rI, \rR\}$.
We consider a continuous-time Markov process,
$(\eta_t)_{t \geq 0}$, where
$\eta_t:=\{\eta_t(x)\}_{x \in \cG}$. 
The transition mechanism is specified by 
two positive parameters 
$\lambda, \mu$ and 
a positive function $c(x, \eta)$ as
\begin{align*}
\bP^{\eta}(\eta_t(x)=\rI, \eta(x)=\rS)
&=\lambda c(x, \eta) t + \ro(t),
\nonumber\\
\bP^{\eta}(\eta_t(x)=\rR, \eta(x)=\rI)
&=\mu t + \ro(t)
\quad \mbox{as $t \to 0$},
\end{align*}
where $\bP^{\eta}$ denotes
the probability in this Markov process
starting from the configuration $\eta$. 
That is, if an individual at point $x$ is in the
susceptible (S) state, it is infected (I)
at rate $\lambda c(x, \eta)$, 
while if it is infected (I), it becomes recovered 
at rate $\mu$. 
Note that once $\eta(x)=\rR$, the state does not change
forever.
We require that only one-variable-change
happens in each transition; that is,
$\bP^{\eta}(\eta_t(x) \not= \eta(x), \eta_t(y) \not= \eta(y))
=\ro(t)$ as $t \to 0$
for each $x, y \in \cG$ with
$x \not=y$ given a configuration $\eta$ 
on $\cG$. 
See \cite{Lig85,Lig99} for the mathematical construction
of interacting particle systems. 

Now we introduce a positive function
$\Psi(n)$, $n \in \N_0$ and specify the function
$c(x, \eta)$ as follows,
\begin{equation}
c(x, \eta)
=\Psi \left(
\sum_{y: |y-x| < 2r} 1_{(\eta(y)=\rI)} \right),
\label{eqn:c1}
\end{equation}
where $1_{(\omega)}$ is the indicator function of
an event $\omega$;
$1_{(\omega)}=1$ if $\omega$ is satisfied and
$1_{(\omega)}=0$ otherwise.
On an infinite percolation cluster $\cG$,
we can discuss the percolation problem
for an infection cluster consisting of
$\rI$-individuals and $\rR$-individuals.
When $\Psi(n)$ is an increasing function of $n$,
the percolation probability
of infection cluster $\Theta^{\rm SIR}$
on $\cG$ is increasing in $\lambda$.
Moreover, a unique critical value $\lambda_{\rm c}$
is defined so that \cite{Gra83,CG85,dST10,TZ10,Zif21}
\begin{align*}
\Theta^{\rm SIR}(\lambda)=0, 
\quad \mbox{if $\lambda \leq \lambda_{\rm c}$},
\nonumber\\
\Theta^{\rm SIR}(\lambda) >0, 
\quad \mbox{if $\lambda > \lambda_{\rm c}$}.
\end{align*}
For each SIR model with a specified $\Psi$, 
it is expected that $\lambda_{\rm c}$ does not
depend on each sample of infinite 
underlying graph $\cG$ and
it is determined only by
the filling factor $\kappa$
of the Boolean percolation model
from which $\cG$ is defined. 

When $\Psi(n)=n$, the infection rate is
just given by the total number of 
infected neighbors multiplied by $\lambda$.
We call this case the
\textit{linear model}.
\section{Numerical Simulations and Results of
the Percolation Problem}
\label{sec:results_percolation}
\subsection{Percolation probability and critical filling factor}
\label{sec:percolation_prob}

\begin{figure}[h]
\includegraphics[width=1\linewidth]{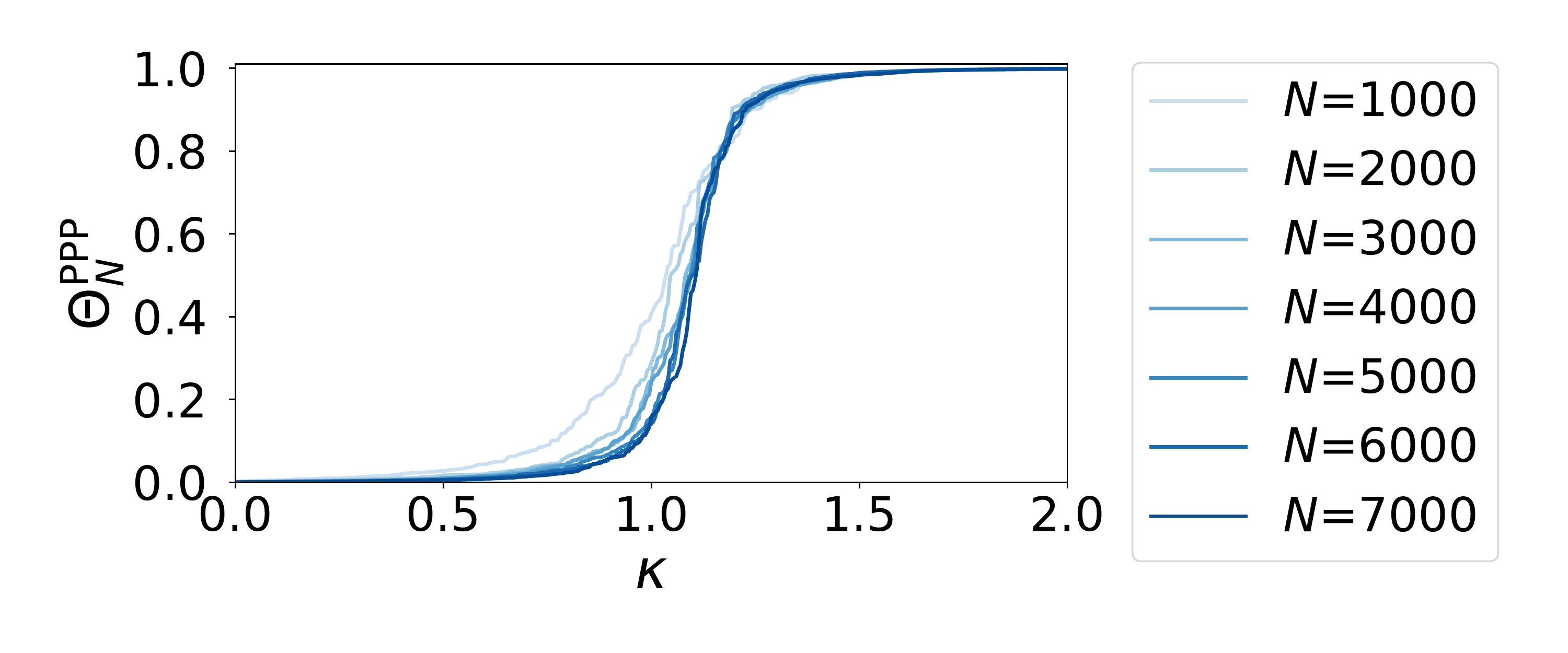}
\includegraphics[width=1\linewidth]{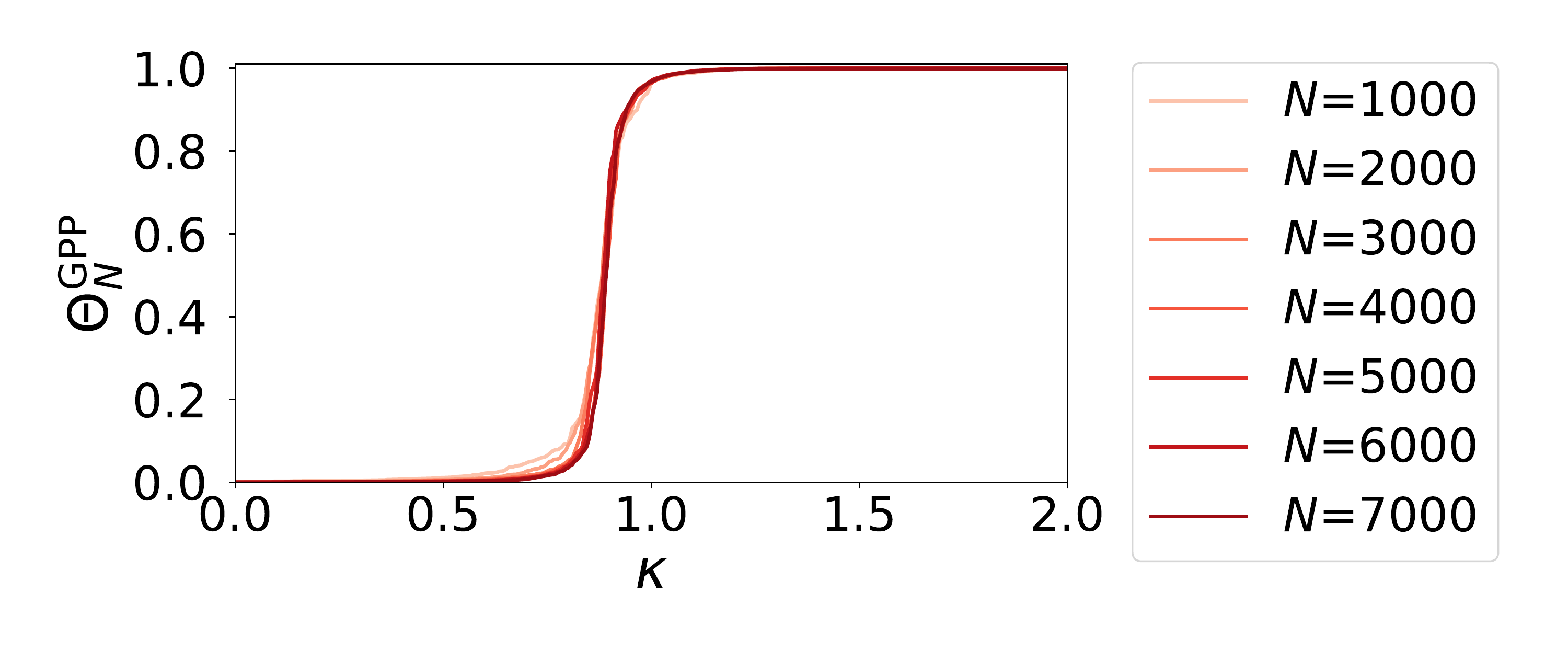}
\caption{
Approximation curves of the percolation probability,
$\Theta^{\rm PPP}_N$ (resp. $\Theta^{\rm GPP}_N$)
are plotted in the top (resp. bottom) figure with respect to
the filling factor $\kappa$ for
several values of $N$.
}
\label{fig:percolation_prob}
\end{figure}

It is easy to generate samples of the PPP with finite number of
points $N$ in the unit square $[0,1]^2$ on $\R^2$,
since it is a uniform point process without any correlation.
The obtained point process is denoted by
$\Xi^{\rm PPP}_{N}$ which has
density $\rho^{\rm PPP}_N=N$.
For the GPP, we have first prepared $L \times L$ random matrices
with $L=2N$, 
whose complex entries are i.i.d.\
and following (\ref{eqn:cNormal}) with 
$\sigma^2=2$; that is, 
the real and the imaginary parts of each entry are
i.i.d.\ real standard Gaussian random variables.
Then we calculated $L=2N$ complex eigenvalues and
plot them on $\C$.
We have confirmed that the eigenvalues
follows the \textit{circle law} \cite{Meh04,For10}.
That is, almost all of them are distributed
in a disk around the origin
with radius $\sqrt{2 N}$, and within the disk, 
except for a very narrow region near the edge of
the disk (i.e., near the circumference), 
their distribution is uniform.
Here we have used only the point distribution in the
central $\sqrt{\pi N} \times \sqrt{\pi N}$ square
and performed a scale change by factor $1/\sqrt{\pi N}$
to obtain a point process $\Xi^{\rm GPP}_N$
on the unit square $[0,1]^2$
having density $\rho^{\rm GPP}_N=N$, 
which approximates the GPP.
In this way we have generated samples of
finite approximations
$\Xi^{\rm PPP}_N$ and $\Xi^{\rm GPP}_N$
with seven different $N$
from $N=1000$ to $N=7000$.

For each finite point processes $\Xi^{\rm PPP}_N$, 
$\Xi^{\rm GPP}_N$ plotted on $[0, 1]^2$, 
we impose the periodic boundary condition.
Then the percolation probability $\Theta$ is
approximated by the ratio 
$\Theta_N$ of the number of points
included in the largest cluster 
to the total number of points $N$ \cite{SA92}.
Figure \ref{fig:percolation_prob} shows the curves of
$\Theta_N$ as functions of the filling factor $\kappa$ 
defined by (\ref{eqn:filling_factor})
for the Boolean percolation models
on $\Xi^{\rm PPP}_N$ in the top
and on $\Xi^{\rm GPP}_N$ in the bottom figure, respectively,
with increasing $N$ from 1000 to 7000
by 1000.
\begin{figure}[ht]
\includegraphics[width=1\linewidth]{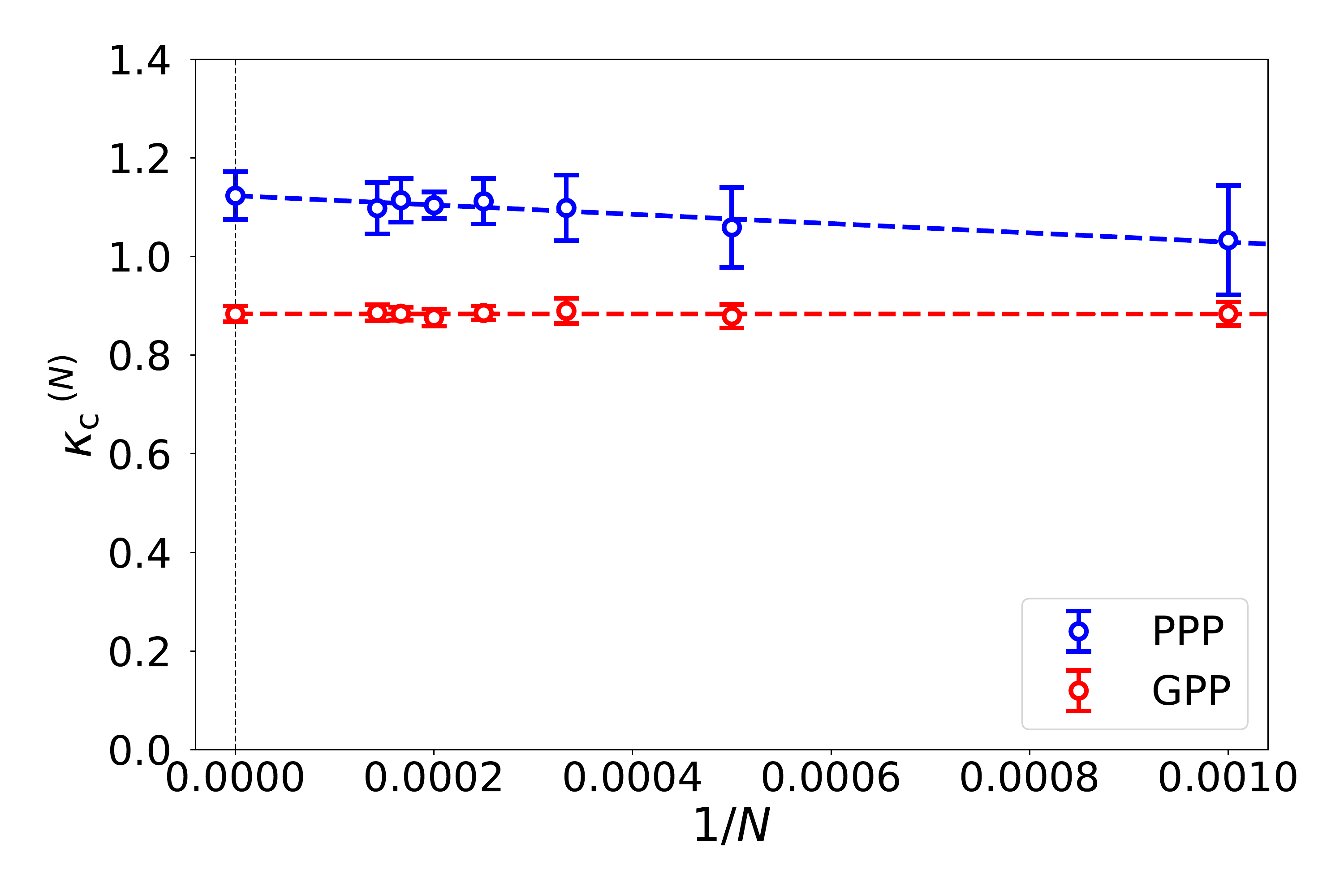}
\caption{
Approximate values of critical filling factor
$\kappa_{\rm c}^{{\rm PPP}(N)}$ 
and $\kappa_{\rm c}^{{\rm GPP}(N)}$ are plotted
versus $1/N$.
The error bars are estimated by ten samples of
finite point processes of each size $N$. 
}
\label{fig:kc_1/N}
\end{figure}
\begin{figure}[!b]
\includegraphics[width=1\linewidth]{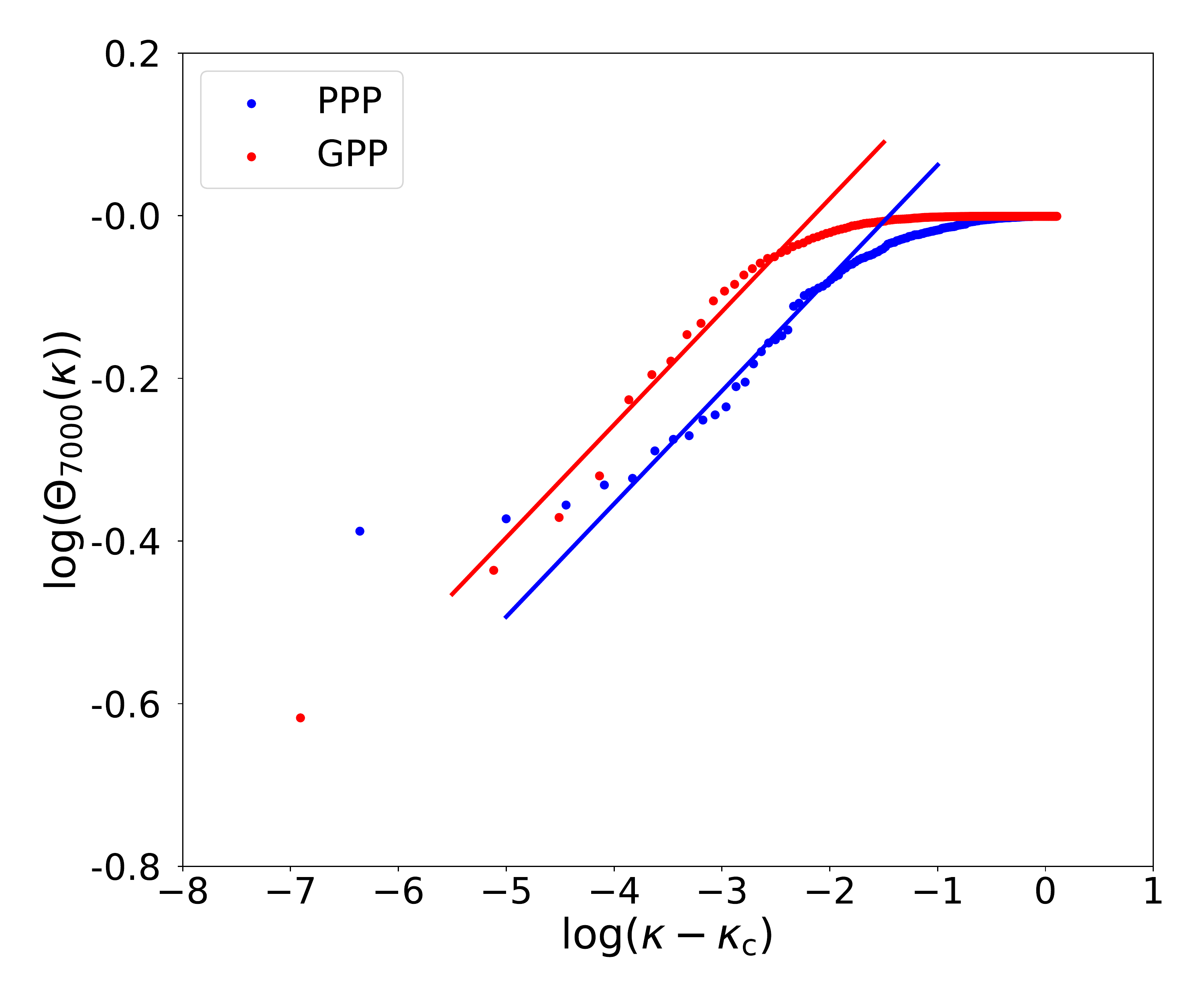}
\caption{%
The least-square linear regression was applied to
the log-log plot of (\ref{eqn:beta1}) with
$\beta=5/36$ for the data of the largest systems with $N=7000$.
The evaluated values of the critical filling factors are
$\kappa_{\rm c}^{\rm PPP}=1.13$ and
$\kappa_{\rm c}^{\rm GPP}=0.894$, respectively.
}
\label{fig:log_log_plot}
\end{figure}

For each curve $\Theta_N$,
an approximate value of critical filling factor,
denoted by $\kappa_{\rm c}^{(N)}$, is
defined by the value of $\kappa$ at which
the numerical value of 
$d \Theta_N(\kappa)/d \kappa$ attains
the maximum.
We plot $\kappa_{\rm c}^{{\rm PPP}(N)}$ and
$\kappa_{\rm c}^{{\rm GPP}(N)}$ 
versus $1/N$ in Fig.~\ref{fig:kc_1/N}.
They behave well as
$\kappa_{\rm c}^{(N)}
\simeq {\rm const.}/N\,+\,\kappa_{\rm c}$
as $N \to \infty$, 
and we obtained the following estimations for 
the critical values,
\begin{align}
\kappa_{\rm c}^{\rm PPP}
&=1.12 \pm 0.05,
\nonumber\\
\kappa_{\rm c}^{\rm GPP}
&=0.884 \pm 0.016.
\label{eqn:kc}
\end{align}

Using the data of the largest systems with $N=7000$,
we performed another method to evaluate $\kappa_{\rm c}$
as follows.
We assumed the critical behavior (\ref{eqn:beta1})
of $\Theta(\kappa)$ with 
the exponent $\beta=5/36$
and applied the least-square linear regression to 
log-log plots of the data
$\Theta_{7000}(\kappa)$ versus
$\kappa-\kappa_{\rm c}$,
where $\kappa_{\rm c}$ is a fitting parameter
and the slope of fitting line 
is fixed to be $5/36$.
Figure \ref{fig:log_log_plot} shows the best fitting,
which gives the estimates,
\begin{equation}
\kappa_{\rm c}^{\rm PPP}=1.13, \quad
\kappa_{\rm c}^{\rm GPP}=0.894.
\label{eqn:kc2}
\end{equation}

The present evaluations
of $\kappa_{\rm c}^{\rm PPP}$
are consistent with the value 
reported in \cite{MM12}
evaluated using efficient algorithms
\cite{QT99,QTZ00,NZ01,QZ07},
$\kappa_{\rm c}^{\rm PPP}=1.12808737(6)$. 
To the best of the knowledge of the present authors,
the above are the first numerical estimations of
the critical filling factor $\kappa_{\rm c}^{\rm GPP}$
of the Boolean percolation model
on the GPP.

\subsection{Largest percolation clusters
as underlying graphs for the SIR models}
\label{sec:largest}

\begin{figure}[ht]
\begin{minipage}{0.45\hsize}
\includegraphics[width=1.2\textwidth]{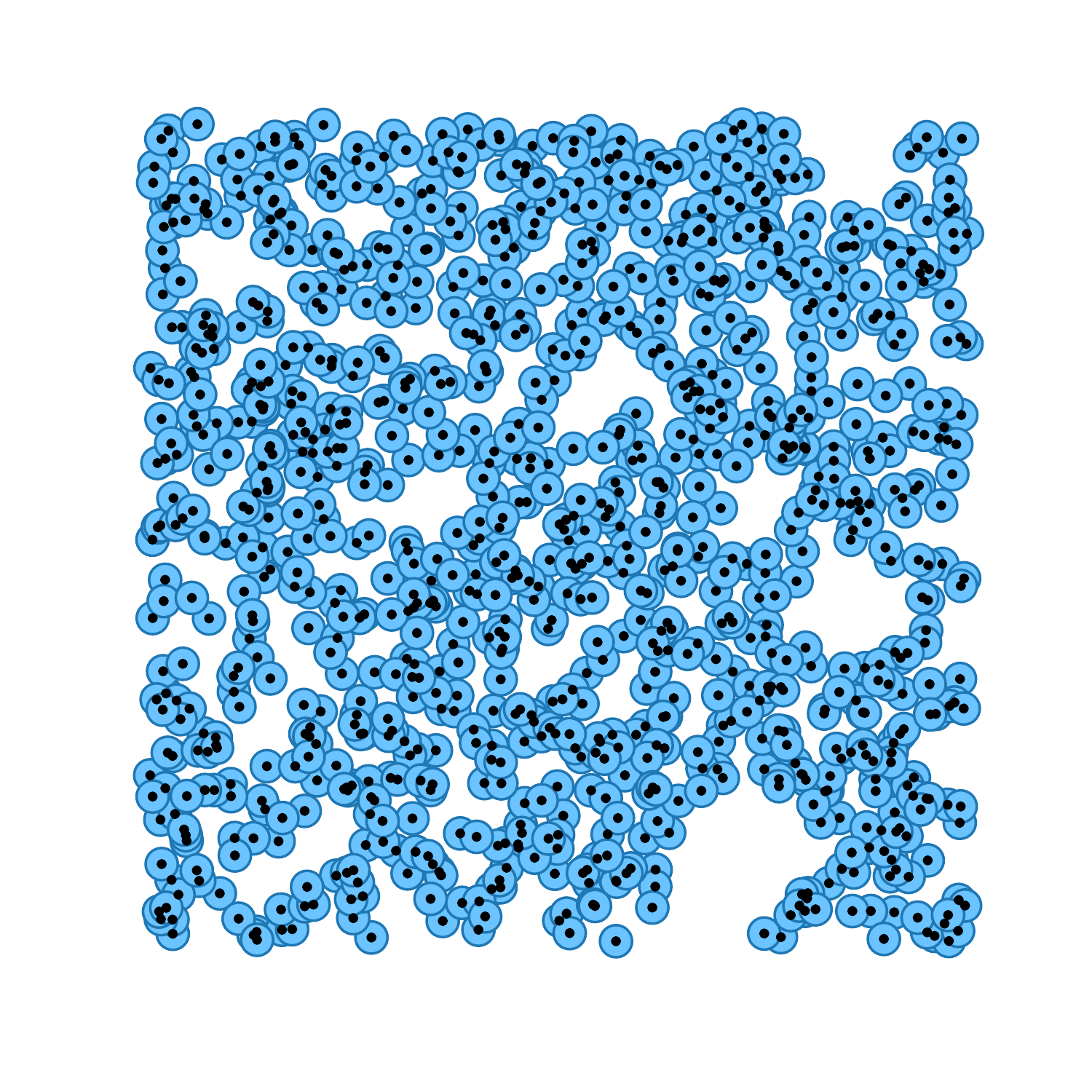}
\end{minipage}
\begin{minipage}{0.45\hsize}
\includegraphics[width=1.2\textwidth]{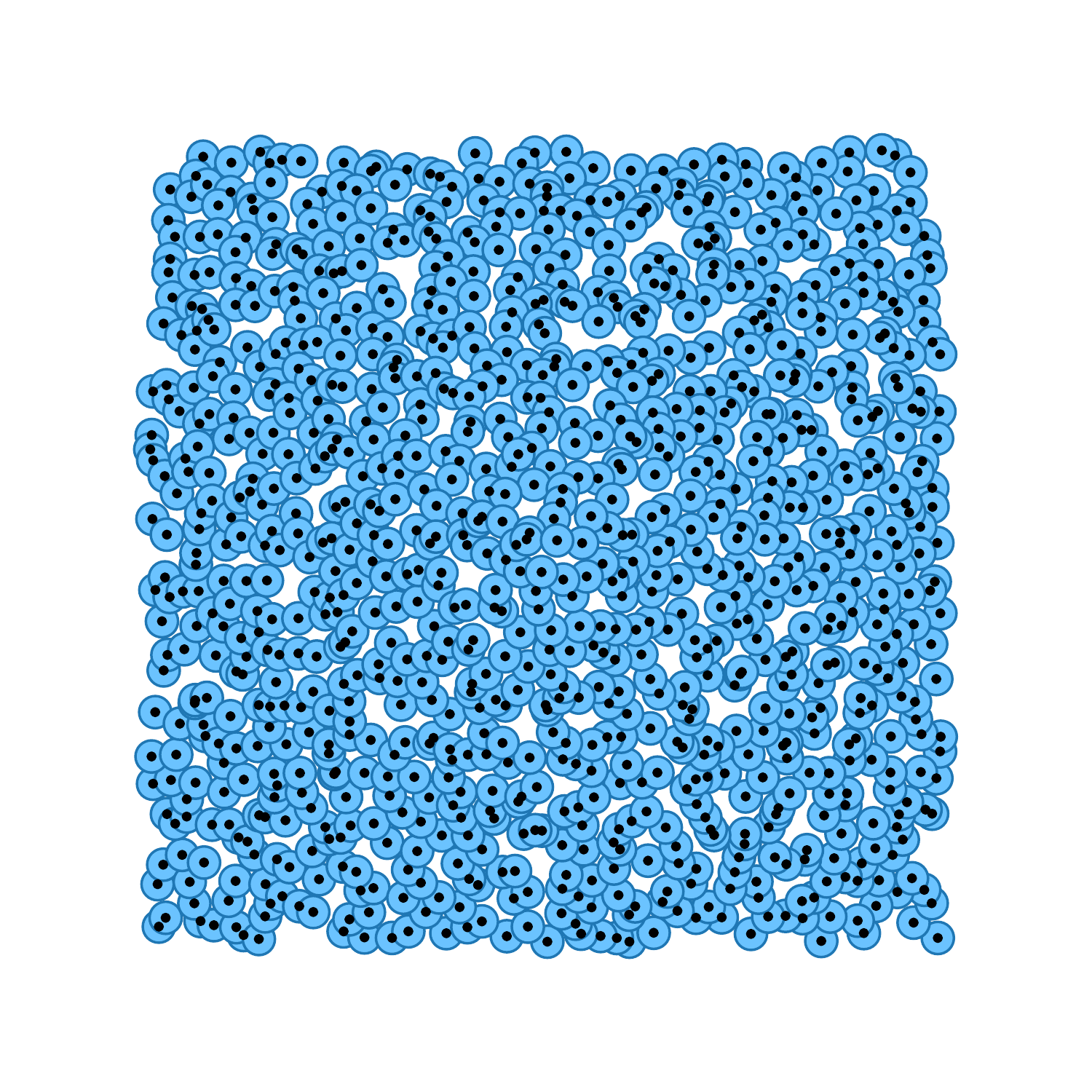}
\end{minipage}
\caption{%
A pair of samples of underlying graphs for 
our contagious epidemic models.
The left graph $\cG^{\rm PPP}_N$ 
(resp. the right graph $\cG^{\rm GPP}_{N}$)
is a finite approximation with $N=1000$ 
of an infinite percolation cluster
$\cG^{\rm PPP}$ (resp. $\cG^{\rm GPP}$).
}
\label{fig:fields}
\end{figure}

From now on we consider the supercritical percolation models.
We fix the filling factor
of the Boolean percolation model as $\kappa=1.3$,
which is greater than the both critical values
$\kappa_{\rm c}^{\rm PPP}$ and
$\kappa_{\rm c}^{\rm GPP}$ evaluated as (\ref{eqn:kc}) and (\ref{eqn:kc2}). 
We pick up only the largest cluster found in the Boolean 
percolation models simulated as explained above.
Figure \ref{fig:fields} shows the typical samples 
for the PPP in the left and the GPP in the right.
In the both systems, 
the total number of disks in the largest cluster is 1000.
We will use such a pair of percolation clusters 
made from point processes
as the underlying graphs $\cG$ of our contagious
epidemic model of the SIR type. 

Although the filling factor $\kappa$ is the same
in the original Boolean percolation models, it is obvious that
these two kinds of $\cG_N$ obtained from them
are different from each other.
The difference can be clearly shown by the distributions
of degree of each point, 
which was already shown by Fig.~\ref{fig:degree_distribution} 
and explained in Section \ref{sec:introduction}. 

\section{Numerical Simulations and Results of
the SIR Models}
\label{sec:results_SIR}
\subsection{The Gillespie algorithm}
\label{sec:Gillespie}
Since time scale is arbitrary in computer simulations,
we can put the recovering rate $\mu \equiv 1$ without
loss of generality.
We consider the SIR model defined in Section \ref{sec:SIR}
on finite underlying graphs $\cG_N$ which were constructed
above from the PPP and the GPP.
If we specify $\lambda$ and $\Psi$, and
a configuration $\eta=\{\eta(x)\}_{\cG_N}$ is given,
a non-negative transition rate $\gamma(x)$ is assigned to each point
$x \in \cG_N$.
We compute the sum
$\gamma=\sum_{x \in \cG_N} \gamma(x)$.
The time $T_1$ at which the configuration is changed
from $\eta$ is assumed to follow the exponential distribution
with the rate $\gamma$;
\[
\bP(T_1 \in d t)
=\gamma e^{-\gamma t} dt, \quad t \geq 0.
\]
At time $T_1$, a site $x \in \cG_N$ is chosen with
probability $\gamma(x)/\gamma$ and we change
the variable $\eta(x)$ according to the transition
having the rate $\gamma(x)$,
and a new configuration $\eta_{T_1}$ is obtained.
Then we repeat the same procedure
on $\eta_{T_1}$ to determine the next time $T_2$
and update the configuration to $\eta_{T_2}$.
This method to simulate continuous Markov process
is known as the Gillespie algorithm \cite{Gil76,Gil77,EC20}. 
(It is also called the {\it n}-fold way algorithm for spin systems \cite{BKL75}.)
The initial configuration $\eta_0$ is given by
the following; one point $x \in \cG_N$ is chosen
at random and put $\eta_0(x)=\rI$ and
$\eta_0(y)=\rS$, $\forall y \not=x, y \in \cG_N$.

\subsection{Critical infection rates}
\label{sec:lambda_c}

First we studied the linear SIR model 
in which we assume 
\begin{equation}
\Psi(n)=n, \quad n \in \N 
\label{eqn:linear}
\end{equation}
for (\ref{eqn:c1}). 
That is, the infection rate is
proportional to the total number of
infected neighbors.
Since $|\cG_N|=N < \infty$, any infection process on $\cG_N$
ceases sooner or later and in
a final configuration 
we have a percolation cluster consisting of
$\rR$-individuals embedded in a field of
$\rS$-individuals. 
For each $\lambda$, we simulated the SIR model
from a single infected individual 
100 times and evaluated the mean ratio
of total number of $\rR$-individuals to
$|\cG_N|=N$ in a final configuration.
We regard this as an approximation
of infection probability $\Theta^{\rm SIR}_N(\lambda)$
with size $N$ \cite{Gra83,CG85,dST10,TZ10,Zif21}.
Increasing the size of system $N$
systematically, we prepare
a series of approximations
$\{\Theta^{\rm SIR}_N(\lambda)\}_{N}$.
Following the procedures similar
to those reported in Section \ref{sec:percolation_prob},
for 
the SIR models
based on the Boolean percolation models with $\kappa=1.3$ we evaluated as 
\begin{align}
\lambda_{\rm c}^{\rm PPP} 
&=4.88 \pm 0.88,
\nonumber\\
\lambda_{\rm c}^{\rm GPP}
&=1.91 \pm 0.02.
\label{eqn:lambda_c}
\end{align}
It will be an important problem to compare the critical infection rates
of the present SIR models with those of the lattice SIR models.
For example, Tom\'e and Ziff reported the value
$\lambda_{\rm c}=4.66571(3)$ for the models on a square lattice \cite{TZ10}.
We should notice, however, that the present SIR models are off-lattice models
and hence $\lambda_{\rm c}$ depends on the filling factor $\kappa$ of
underlying point processes. The above values (\ref{eqn:lambda_c}) are for the SIR models on the
PPP and the GPP, respectively, both with a specified value $\kappa=1.3$.
In a forthcoming paper, we will report a systematic study of the
dependence of $\lambda_{\rm c}$ on $\kappa$ both for the PPP-based
and the GPP-based SIR models \cite{MK2}.
Comparison with the lattice SIR models
will be discussed there.

\subsection{Cumulative number of infected 
individuals in subcritical regime}
\label{sec:NI}

In the following, we study the SIR models
when the infectivity $\lambda$ 
is much smaller than $\lambda_{\rm c}$
both for the PPP-based model
and for the GPP-based model;
for instance, $\lambda \lesssim 1.4$ as shown 
in Fig.~\ref{fig:I_vs_lambda_quadratic}.
Therefore, the finite-size effect is
irrelevant and $\cG_N$ with $N = 1000$
shown by Fig.~\ref{fig:fields} will approximate
an infinite percolation cluster $\cG$ very well
in our study.
We prepared ten distinct samples
$\cG^{\rm PPP}_N$
and $\cG^{\rm GPP}_N$ with $N = 1000$.
From now on, we write
a pair of such samples as
$\cG^{\rm PPP}$ and $\cG^{\rm GPP}$
omitting the subscript $N$
and call them simply the
PPP- and GPP-underlying graphs.

\begin{figure}[h!]
\includegraphics[width=1\linewidth]{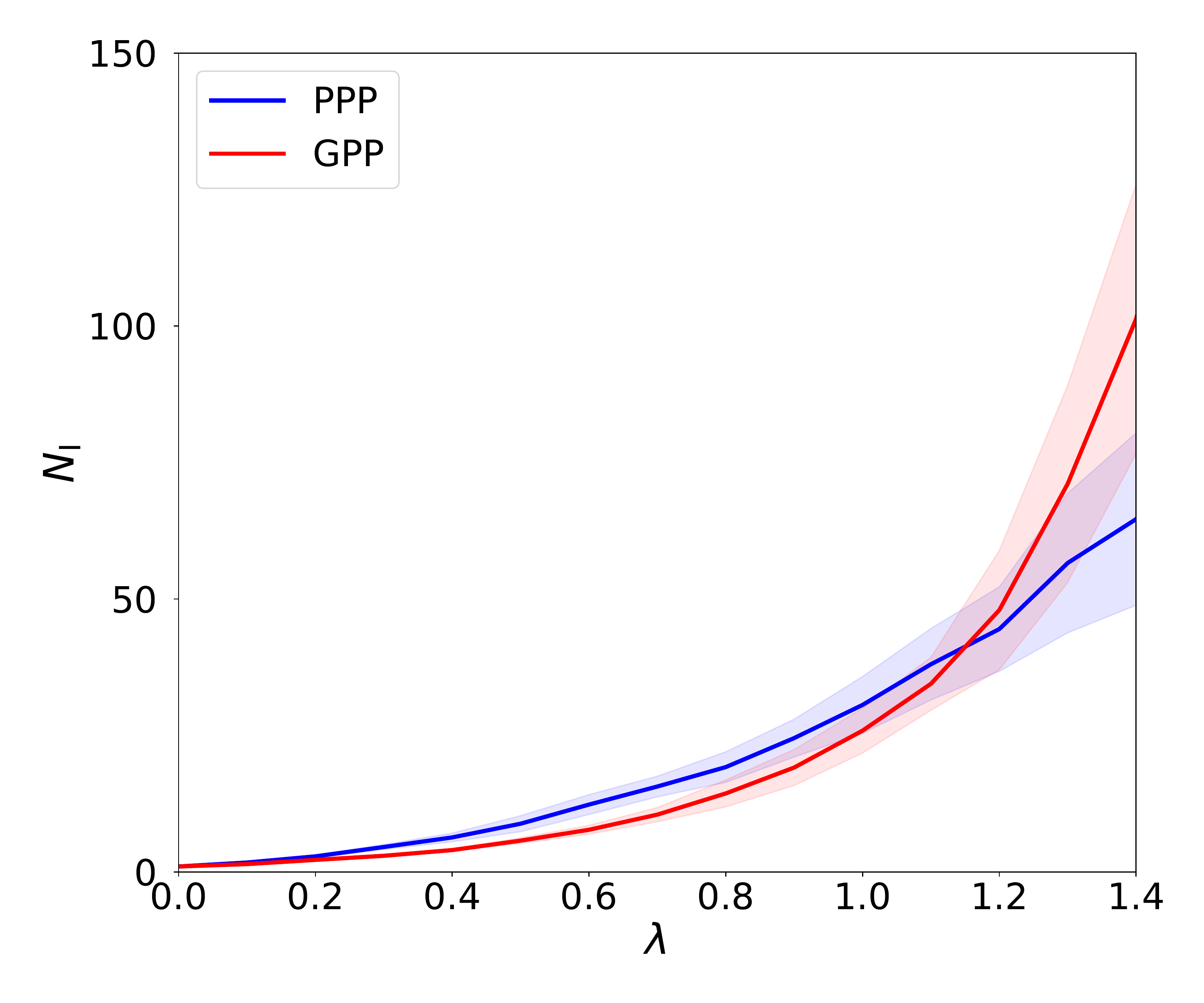}
\caption{%
For the SIR model with (\ref{eqn:linear}) 
the quenched averages $\bra \cN_{\rI} \ket$
of the cumulative numbers of infected individuals
(\ref{eqn:NI}) are
shown versus $\lambda$ for
the PPP-based model by a blue curve
and for the GPP-based model by a red curve.
The standard deviations in the ten samples are shown by
shaded strips around the curves.
We find a crossing point of these two curves
at $\lambda_* \simeq 1.2$.
}
\label{fig:I_vs_lambda_linear}
\end{figure}

In order to measure the whole size of an infection
process starting from one infected individual, 
we define the cumulative number of
individuals who were infected during the process,
\begin{equation}
\cN_{\rI}:=1 + \sum_{j \geq 1} \sum_{x \in \cG} 
1_{(\eta_{T_j}(x)=\rI, \eta_{T_{j-1}}(x)=\rS)},
\label{eqn:NI}
\end{equation}
where $0=:T_0 < T_1 < T_2 < \cdots$ is 
a sequence of times at each of which 
a transition with one-variable-change occurred.
Notice that $\cN_{\rI}$ is equal to the
total number of $\rR$-individuals in the
final configuration.
With a given value of $\lambda$ the SIR process
was simulated 100 times
and the average of $\cN_{\rI}$ was calculated.
Then we evaluate the mean $\bra \cN_{\rI} \ket$
and the standard deviation
over ten distinct samples of $\cG$.
Since underlying graph $\cG$ is 
generated from a random point process $\Xi$
using the notion of percolation cluster,
it is regarded as a random environment, 
and the present SIR model is a 
\textit{random process
in random environment}.
If we use the notations used in
Section {\ref{sec:models}},
the averaging over 100 runs of the SIR model 
following the Gillespie algorithm 
is a statistical procedure with respect to 
the probability law $\bP$ given 
one sample of $\cG$, 
and this will be said to be
a \textit{quenched averaging}.
The averaging over ten samples of $\cG$
is regarded as a procedure 
with respect to the probability law $\rP$
which governs the underlying point processes. 
The results are shown in Fig.~\ref{fig:I_vs_lambda_linear}. 
We find a special value $\lambda_* \simeq 1.2$
such that
$\bra \cN_{\rI}^{\rm PPP} \ket > \bra \cN_{\rI}^{\rm GPP} \ket$
if $0 < \lambda < \lambda_*$,
while 
$\bra \cN_{\rI}^{\rm PPP} \ket < \bra \cN_{\rI}^{\rm GPP} \ket$
if $\lambda > \lambda_*$.

\begin{figure}[ht]
\includegraphics[width=1\linewidth]{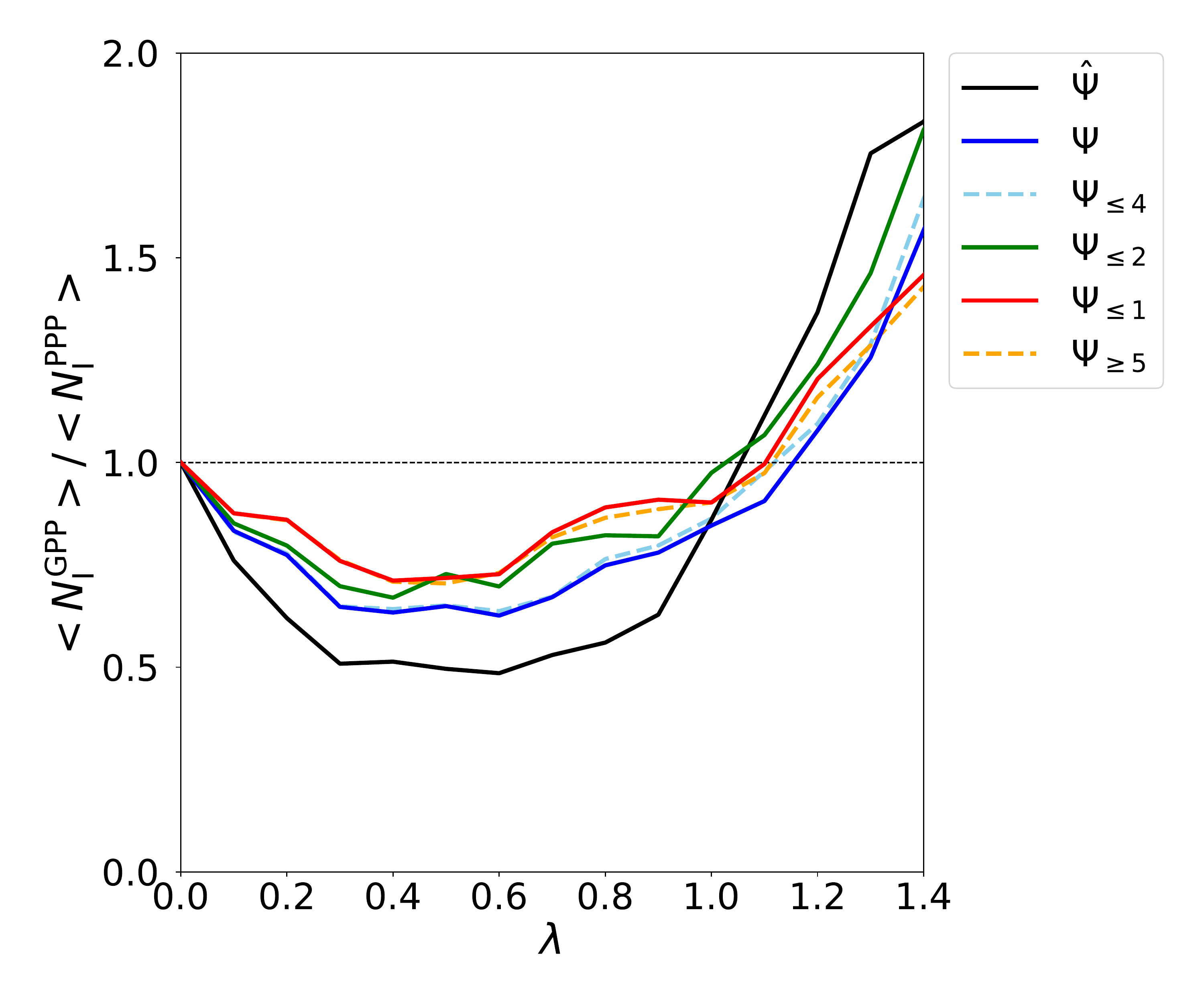}
\caption{%
Ratios $\bra \cN^{\rm GPP}_{\rI} \ket/
\bra \cN^{\rm PPP}_{\rI} \ket$
are plotted versus $\lambda$ for
several models with different choices of
$\Psi$. 
}
\label{fig:GPP_PPP_ratio}
\end{figure}

We thought that the difference between
$\bra \cN_{\rI}^{\rm PPP} \ket$ and
$\bra \cN_{\rI}^{\rm GPP} \ket$ can be
simply attribute to the difference in the
distributions of degree shown by
Fig.~\ref{fig:degree_distribution} 
in Section \ref{sec:introduction}.
In order to verify this naive conjecture, we considered
the following modifications of (\ref{eqn:linear}),
\begin{align*}
\Psi_{\leq 1}(n) &= 1, \quad \forall n \geq 1,
\nonumber\\
\Psi_{\leq 2}(n) &= 
\begin{cases}
n, & \quad (n=1, 2),
\cr
1, & \quad (n \geq 3),
\end{cases}
\nonumber\\
\Psi_{\leq 4}(n) &= 
\begin{cases}
n, & \quad (n=1, 2, 3, 4),
\cr
1, & \quad (n \geq 5),
\end{cases}
\nonumber\\
\Psi_{\geq 5}(n) &= 
\begin{cases}
1, & \quad (n=1, 2, 3, 4),
\cr
n, & \quad (n \geq 5).
\end{cases}
\end{align*}
The ratios
$\bra \cN_{\rI}^{\rm GPP} \ket
/\bra \cN_{\rI}^{\rm PPP} \ket$
are plotted versus $\lambda$
in Fig.~\ref{fig:GPP_PPP_ratio} 
for these different models
as well as for the original linear model
(\ref{eqn:linear}) 
and for the quadratic model
with $\widehat{\Psi}(n)=n^2$.
We found the similarities between
the models with $\Psi_{\leq 1}$ and
with $\Psi_{\geq 5}$, 
and between 
the model with $\Psi_{\leq 4}$ and 
the original linear model (\ref{eqn:linear}).
These observations imply that
the cases with $n \geq 5$ do not contribute
to the present phenomenon.
So we reconsidered the original linear model
more carefully.

\begin{figure}[h!]
\includegraphics[width=1\linewidth]{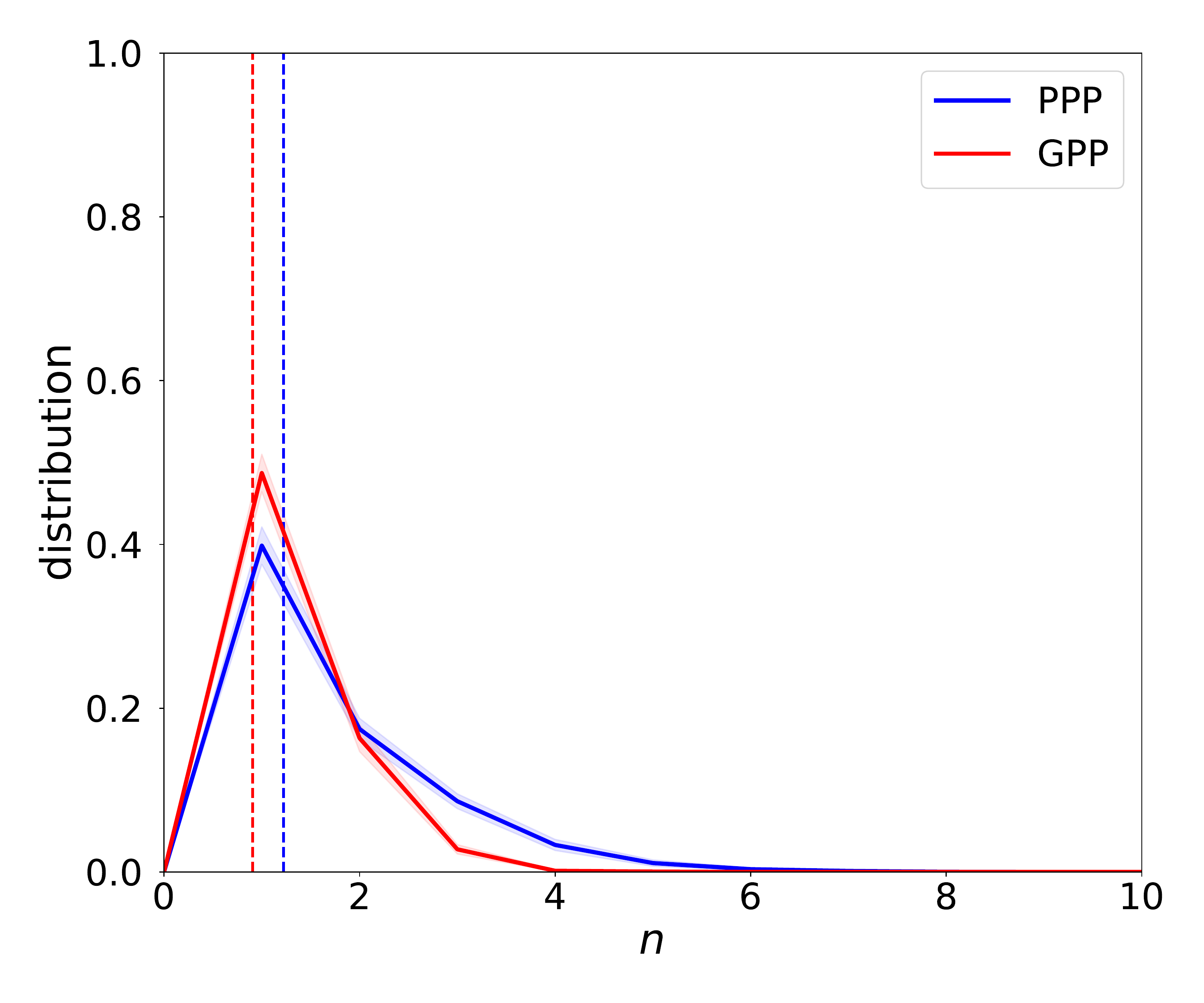}
\caption{%
Distributions of $n$ are shown 
for the PPP-based and the GPP-based
SIR model with (\ref{eqn:linear}), when $\lambda=0.5$. 
Solid lines show the quenched averages 
of ten distinct samples of $\cG$ 
and shaded strips show the standard deviations of them.
The means of $n$ in the PPP-based
model and the GPP-based model
are 1.22 and 0.904, respectively.
}
\label{fig:distribution_n}
\end{figure}

In each realization of process, we traced 
a time-sequence of points at each of which 
a transition $\rS \to \rI$ was taken place,
and counted the number of infected individuals
in the neighborhood of each point.
We obtained the distribution of the number of
infected neighbors $n$ using the data of 100 runs
on each underlying graph $\cG$, 
and then we calculated the quenched average
of distributions
over ten samples of $\cG$.
Figure \ref{fig:distribution_n} shows the
distributions of $n$
when $\lambda =0.5$ 
for the PPP-based and the GPP-based models.

Compared with the distributions of
degree in underlying graphs
shown by Fig.~\ref{fig:degree_distribution},
any contribution from
\textit{hubs} with a large number of neighbors
\cite{Bar16} are truncated in the `effective
network' for the present SIR models.
The difference of distributions
of $n$ between the PPP-based model
and the GPP-based model seems
to be very small in 
Fig.~\ref{fig:distribution_n},
but this difference definitely causes 
the discrepancy between the curve
for the model with $\Psi_{\leq 1}$ (to which
the curve for the model with $\Psi_{\geq 5}$ is similar)
and that with the original linear function
(\ref{eqn:linear})
(to which the curve with $\Psi_{\leq 4}$
is similar) shown in 
Fig.~\ref{fig:GPP_PPP_ratio}. 
We should notice that 
even in the model with $\Psi_{\leq 1}$, 
which has no dependence on $n \geq 1$, 
Fig.~\ref{fig:GPP_PPP_ratio} shows
the suppression 
$\bra \cN_{\rI}^{\rm GPP} \ket
/\bra \cN_{\rI}^{\rm PPP} \ket < 1$
for $\lambda < \lambda_*$.
This effect will be attributed
to \textit{blobs} in underlying percolation
cluster $\cG$ \cite{SA92}.
A blob is constructed on clumping of points
in the PPP caused by the Poissonian 
large fluctuation.
And it is observed as an infection cluster
which seems to be compact 
but included relatively large number
of infected individuals.
The comparison to the GPP-based model
was already demonstrated by
Fig.~\ref{fig:infection_clusters} in Section \ref{sec:introduction}.

As shown by Fig.~\ref{fig:I_vs_lambda_quadratic}
and Fig.~\ref{fig:GPP_PPP_ratio},
the suppression of infection clusters
in the GPP-based model
is much enhanced, when
the contagion is described 
by the `quadratic SIR model'
with $\widehat{\Psi}(n)=n^2$, $n \in \N$.

\section{Concluding Remarks}
\label{sec:concluding_remarks}

Among a variety of continuum percolation 
models (see, for instance, Chapter 8 of \cite{MR96}
and Chapter 8 of \cite{BR06})
we have focused on the Boolean
model with same-radius disks
(the standard Gilbert disk model \cite{Gilb61})
on $\R^2$ in this paper.
Even in this simplest model,
many works have been done 
in the case that the model is
\textit{driven by} the uniform PPP.
On the other hand,
the GPP and other determinantal point processes
(DPPs) have been extensively studied
in random matrix theory \cite{Meh04,For10,Kat15}
and the hyperuniformity of such repelling point processes
has attracted much attention in many research fields
\cite{Tor18,MKS21}.
Recently comparison of 
clustering properties between
the PPP and non-Poisson point processes 
is mathematically studied \cite{BY13,BY14}. 
The continuum percolation models
driven by DPPs are also studied in probability
theory \cite{GKP16}.
In the present paper 
we reported a numerical study on
the Boolean percolation models
defined on the PPP and the GPP.
The purpose of the comparison
between these two kinds of continuum percolation
models here is to construct the SIR models
of infection processes on them
and to clarify differences in 
stochastic properties 
of infection clusters between the
PPP-based and the GPP-based models.
We are interested in hierarchical 
modeling such that
the underlying graph $\cG$ is formed using
a random point process
and a notion of continuum percolation,
and then a time-dependent system is
constructed on $\cG$.

We list out remarks on open problems
and future directions.

\begin{enumerate}
\item
We reported the numerical evaluation of
the critical filling factor
$\kappa_{\rm c}^{\rm GPP}$ for the
Boolean percolation model on $\R^2$
defined on the GPP.
We can find a series of papers
on numerical evaluations of $\kappa_{\rm c}^{\rm PPP}$;
see \cite{QT99,QTZ00,QZ07,MM12} and 
references therein.
Further study on more precise enumeration of
$\kappa_{\rm c}^{\rm GPP}$ should be done.

\item
Our estimations of the critical
infection rates $\lambda_{\rm c}^{\rm PPP}(\kappa)$
and $\lambda_{\rm c}^{\rm GPP}(\kappa)$
of the SIR models
are preliminary and 
they are given only for a special value
$\kappa=1.3$
of the filling factor in the
underlying Boolean percolation models.
Since we have been interested in infection processes
with $\lambda$ which is much smaller than
$\lambda_{\rm c}$ in the present work,
we did not need precise values of
$\lambda_{\rm c}$.
Systematic study on the
critical value $\lambda_{\rm c}(\kappa)$
as a function of $\kappa$
as well as on the critical phenomena
\cite{Gra83,CG85,Hin00,VFM09,dST10,TZ10,Zif21}
observed in the vicinity of 
the critical line $\lambda=\lambda_{\rm c}(\kappa)$,
$\kappa >0$ is required \cite{MK2}. 

\item 
The present numerical results
show the strict inequalities
$\kappa_{\rm c}^{\rm GPP} < \kappa_{\rm c}^{\rm PPP}$
and $\lambda_{\rm c}^{\rm GPP} < \lambda_{\rm c}^{\rm PPP}$.
They suggest that covering processes
and spreading phenomena can be achieved
more efficiently in the GPP-based models
than the PPP-based models.
We find interesting examples
of this fact in the work on
modeling and analysis of cellular networks
\cite{MS14,LBDA15,MS16}.
In the present paper, however, 
we have reported an opposite tendency such that
emergence of infection clusters in
contagious processes is suppressed
in the GPP-based model
compared to the PPP-based model,
if the infectivity $\lambda$ is less than
a special value denoted by $\lambda_*$.
We are planning to proceed the study
reported in Section \ref{sec:NI}
in order to understand the mechanism
determining the values of $\lambda_*$.
Moreover, effect of nonlinearity of
the function $\Psi(n), n \in \N$ on this
phenomenon should be clarified in the future study \cite{MK2}.

\item
As well-defined random point processes
with positive correlations,
\textit{permanental (Boson) point processes}
have been studied in probability theory
\cite{ST03a,ST03b,TI06,TI07,HKPV09,AFY19}.
Infection models on permanental point processes
shall be studied and compared with
the present SIR models on the 
Poisson point process and the Ginibre point process.
\end{enumerate}

\section*{CRediT authorship contribution statement}
\textbf{Machiko Katori:}
Conceptualization, Methodology, Analytical results, Numerical results, Writing.
\textbf{Makoto Katori:}
Conceptualization, Methodology, Writing.

\section*{Acknowledgements}
Machiko K. wishes to thank T. J. Kobayashi for carefully reading of the manuscript and useful comments.
Makoto K. would like to express his gratitude to John W. Essam for a stimulating communication which motivated the present work.
The present authors thank the anonymous referees for valuable comments which are very useful to improve the present paper and to perform further study on this subject.
Machiko K. was supported by the ANRI Fellowship and International Graduate Program of Innovation for Intelligent World (IIW) of The University of Tokyo.
Makoto K. was supported by
the Grant-in-Aid for Scientific Research (C) (No.19K03674),
(B) (No.18H01124),
(S) (No.16H06338),
and
(A) (No.21H04432) 
of Japan Society for the Promotion of Science.

 \bibliographystyle{bst/elsarticle-num} 
 \bibliography{final/tex/MK2-reference-re.bib}

\begin{thebibliography}{10}
\expandafter\ifx\csname url\endcsname\relax
  \def\url#1{\texttt{#1}}\fi
\expandafter\ifx\csname urlprefix\endcsname\relax\def\urlprefix{URL }\fi
\expandafter\ifx\csname href\endcsname\relax
  \def\href#1#2{#2} \def\path#1{#1}\fi

\bibitem{KM27}
W.~O. Kermack, A.~G. McKendrick,
  \href{https://royalsocietypublishing.org/doi/abs/10.1098/rspa.1927.0118
  https://royalsocietypublishing.org/doi/10.1098/rspa.1927.0118}{{A
  contribution to the mathematical theory of epidemics}}, Proc. R. Soc. London.
  Ser. A, Contain. Pap. a Math. Phys. Character 115~(772) (1927) 700--721.
\newblock \href {https://doi.org/10.1098/rspa.1927.0118}
  {\path{doi:10.1098/rspa.1927.0118}}.

\bibitem{Bai53}
N.~T.~J. Bailey,
  \href{https://www.jstor.org/stable/2333107?origin=crossref}{{The total size
  of a general stochastic epidemic}}, Biometrika 40~(1/2) (1953) 177.
\newblock \href {https://doi.org/10.2307/2333107} {\path{doi:10.2307/2333107}}.

\bibitem{Bai57}
N.~T.~J. Bailey, The Mathematical Theory of Epidemics, Hafner, New York, 1957.

\bibitem{Lig85}
T.~M. Liggett, \href{http://link.springer.com/10.1007/b138374}{{Interacting
  Particle Systems}}, Classics in Mathematics, Springer, Berlin, Heidelberg,
  1985.
\newblock \href {https://doi.org/10.1007/b138374} {\path{doi:10.1007/b138374}}.

\bibitem{AM91}
R.~M. Anderson, R.~M. May, {Infectious Diseases of Humans: Dynamics and
  Control.}, Oxford University Press, Oxford, UK, 1991.

\bibitem{Lig99}
T.~M. Liggett,
  \href{http://link.springer.com/10.1007/978-3-662-03990-8}{{Stochastic
  Interacting Systems: Contact, Voter and Exclusion Processes}}, Vol. 324 of
  Grundlehren der mathematischen Wissenschaften, Springer, Berlin, Heidelberg,
  1999.
\newblock \href {https://doi.org/10.1007/978-3-662-03990-8}
  {\path{doi:10.1007/978-3-662-03990-8}}.

\bibitem{DH00}
O.~Diekmann, J.~A.~P. Heesterbeek, {Mathematical Epidemiology of Infectious
  Diseases: Model Building, Analysis and Interpretation}, John Wiley and Sons,
  Chichester, 2000.

\bibitem{CHBC09}
G.~Chowell, J.~M. Hyman, L.~M.~A. Bettencourt, C.~Castillo-Chavez (Eds.),
  \href{http://link.springer.com/10.1007/978-90-481-2313-1}{{Mathematical and
  Statistical Estimation Approaches in Epidemiology}}, Springer Netherlands,
  Dordrecht, 2009.
\newblock \href {https://doi.org/10.1007/978-90-481-2313-1}
  {\path{doi:10.1007/978-90-481-2313-1}}.

\bibitem{Bar16}
A.-L. Barab{\'{a}}si, {Network Science}, Cambridge University Press, Cambridge,
  2016.

\bibitem{Gra83}
P.~Grassberger,
  \href{https://linkinghub.elsevier.com/retrieve/pii/0025556482900360}{{On the
  critical behavior of the general epidemic process and dynamical
  percolation}}, Math. Biosci. 63~(2) (1983) 157--172.
\newblock \href {https://doi.org/10.1016/0025-5564(82)90036-0}
  {\path{doi:10.1016/0025-5564(82)90036-0}}.

\bibitem{dST10}
D.~R. de~Souza, T.~Tom{\'{e}},
  \href{https://linkinghub.elsevier.com/retrieve/pii/S0378437109009091}{{Stochastic
  lattice gas model describing the dynamics of the SIRS epidemic process}},
  Phys. A Stat. Mech. its Appl. 389~(5) (2010) 1142--1150.
\newblock \href {https://doi.org/10.1016/j.physa.2009.10.039}
  {\path{doi:10.1016/j.physa.2009.10.039}}.

\bibitem{TZ10}
T.~Tom{\'{e}}, R.~M. Ziff,
  \href{https://link.aps.org/doi/10.1103/PhysRevE.82.051921}{{Critical behavior
  of the susceptible-infected-recovered model on a square lattice}}, Phys. Rev.
  E 82~(5) (2010) 051921.
\newblock \href {https://doi.org/10.1103/PhysRevE.82.051921}
  {\path{doi:10.1103/PhysRevE.82.051921}}.

\bibitem{SMDHB20}
S.~Saha, A.~Mishra, S.~K. Dana, C.~Hens, N.~Bairagi,
  \href{https://link.aps.org/doi/10.1103/PhysRevE.102.052307}{{Infection
  spreading and recovery in a square lattice}}, Phys. Rev. E 102~(5) (2020)
  052307.
\newblock \href {https://doi.org/10.1103/PhysRevE.102.052307}
  {\path{doi:10.1103/PhysRevE.102.052307}}.

\bibitem{SAAMF20}
G.~Santos, T.~Alves, G.~Alves, A.~Macedo-Filho, R.~Ferreira,
  \href{https://doi.org/10.1016/j.physleta.2019.126063
  https://linkinghub.elsevier.com/retrieve/pii/S0375960119309533}{{Epidemic
  outbreaks on two-dimensional quasiperiodic lattices}}, Phys. Lett. A 384~(2)
  (2020) 126063.
\newblock \href {https://doi.org/10.1016/j.physleta.2019.126063}
  {\path{doi:10.1016/j.physleta.2019.126063}}.

\bibitem{Zif21}
R.~M. Ziff,
  \href{https://linkinghub.elsevier.com/retrieve/pii/S0378437120310219}{{Percolation
  and the pandemic}}, Phys. A Stat. Mech. its Appl. 568 (2021) 125723.
\newblock \href {https://doi.org/10.1016/j.physa.2020.125723}
  {\path{doi:10.1016/j.physa.2020.125723}}.

\bibitem{DVJ03}
D.~Daley, D.~Vere-Jones, \href{http://link.springer.com/10.1007/b97277}{{An
  Introduction to the Theory of Point Processes}}, 2nd Edition, Probability and
  its Applications, Springer-Verlag, New York, 2003.
\newblock \href {https://doi.org/10.1007/b97277} {\path{doi:10.1007/b97277}}.

\bibitem{Gin65}
J.~Ginibre, \href{https://doi.org/10.1063/1.1704292
  http://aip.scitation.org/doi/10.1063/1.1704292}{{Statistical ensembles of
  complex, quaternion, and real matrices}}, J. Math. Phys. 6~(3) (1965)
  440--449.
\newblock \href {https://doi.org/10.1063/1.1704292}
  {\path{doi:10.1063/1.1704292}}.

\bibitem{Meh04}
M.~L. Mehta,
  \href{https://linkinghub.elsevier.com/retrieve/pii/C20090222975}{{Random
  Matrices}}, 3rd Edition, Elsevier, Amsterdam, 2004.
\newblock \href {https://doi.org/10.1016/C2009-0-22297-5}
  {\path{doi:10.1016/C2009-0-22297-5}}.

\bibitem{For10}
P.~J. Forrester,
  \href{https://www.degruyter.com/document/doi/10.1515/9781400835416/html}{{Log-Gases
  and Random Matrices}}, Princeton University Press, Princeton, 2010.
\newblock \href {https://doi.org/10.1515/9781400835416}
  {\path{doi:10.1515/9781400835416}}.

\bibitem{Sos00}
A.~Soshnikov, \href{https://doi.org/10.1070/rm2000v055n05abeh000321
  http://stacks.iop.org/0036-0279/55/i=5/a=R02?key=crossref.e0bbd04630a1991ba0aee89bdcd2ed62}{{Determinantal
  random point fields}}, Russ. Math. Surv. 55~(5) (2000) 923--975.
\newblock \href {https://doi.org/10.1070/RM2000v055n05ABEH000321}
  {\path{doi:10.1070/RM2000v055n05ABEH000321}}.

\bibitem{ST03a}
T.~Shirai, Y.~Takahashi,
  \href{https://linkinghub.elsevier.com/retrieve/pii/S002212360300171X}{{Random
  point fields associated with certain Fredholm determinants I: fermion,
  Poisson and boson point processes}}, J. Funct. Anal. 205~(2) (2003) 414--463.
\newblock \href {https://doi.org/10.1016/S0022-1236(03)00171-X}
  {\path{doi:10.1016/S0022-1236(03)00171-X}}.

\bibitem{ST03b}
T.~Shirai, Y.~Takahashi,
  \href{http://projecteuclid.org/euclid.aop/1055425789}{{Random point fields
  associated with certain Fredholm determinants II: fermion shifts and their
  ergodic and Gibbs properties}}, Ann. Probab. 31~(3) (2003) 1533--1564.
\newblock \href {https://doi.org/10.1214/aop/1055425789}
  {\path{doi:10.1214/aop/1055425789}}.

\bibitem{Shi06}
T.~Shirai, \href{http://link.springer.com/10.1007/s10955-006-9026-x}{{Large
  deviations for the fermion point process associated with the exponential
  kernel}}, J. Stat. Phys. 123~(3) (2006) 615--629.
\newblock \href {https://doi.org/10.1007/s10955-006-9026-x}
  {\path{doi:10.1007/s10955-006-9026-x}}.

\bibitem{HKPV09}
J.~Hough, M.~Krishnapur, Y.~Peres, B.~Vir{\'{a}}g,
  \href{http://www.ams.org/ulect/051}{{Zeros of Gaussian Analytic Functions and
  Determinantal Point Processes}}, Vol.~51 of University Lecture Series,
  American Mathematical Society, Providence, Rhode Island, 2009.
\newblock \href {https://doi.org/10.1090/ulect/051}
  {\path{doi:10.1090/ulect/051}}.

\bibitem{Kat15}
M.~Katori, \href{http://link.springer.com/10.1007/978-981-10-0275-5}{{Bessel
  Processes, Schramm–Loewner Evolution, and the Dyson Model}}, Vol.~11 of
  SpringerBriefs in Mathematical Physics, Springer, Singapore, 2016.
\newblock \href {https://doi.org/10.1007/978-981-10-0275-5}
  {\path{doi:10.1007/978-981-10-0275-5}}.

\bibitem{Tor18}
S.~Torquato,
  \href{https://www.sciencedirect.com/science/article/pii/S037015731830036X
  https://linkinghub.elsevier.com/retrieve/pii/S037015731830036X}{{Hyperuniform
  states of matter}}, Phys. Rep. 745 (2018) 1--95.
\newblock \href {https://doi.org/10.1016/j.physrep.2018.03.001}
  {\path{doi:10.1016/j.physrep.2018.03.001}}.

\bibitem{MKS21}
T.~Matsui, M.~Katori, T.~Shirai,
  \href{http://iopscience.iop.org/article/10.1088/1751-8121/abecaa
  https://iopscience.iop.org/article/10.1088/1751-8121/abecaa
  http://arxiv.org/abs/2012.10585}{{Local number variances and hyperuniformity
  of the Heisenberg family of determinantal point processes}}, J. Phys. A Math.
  Theor. 54~(16) (2021) 165201.
\newblock \href {https://doi.org/10.1088/1751-8121/abecaa}
  {\path{doi:10.1088/1751-8121/abecaa}}.

\bibitem{Gilb61}
E.~N. Gilbert, \href{http://www.jstor.org/stable/2098879}{{Random plane
  networks}}, J. Soc. Ind. Appl. Math. 9~(4) (1961) 533--543.

\bibitem{MR96}
R.~Meester, R.~Roy,
  \href{https://www.cambridge.org/core/product/identifier/9780511895357/type/book}{{Continuum
  Percolation}}, Cambridge Tracts in Mathematics, Cambridge University Press,
  Cambridge, 1996.
\newblock \href {https://doi.org/10.1017/CBO9780511895357}
  {\path{doi:10.1017/CBO9780511895357}}.

\bibitem{BR06}
B.~Bollobas, O.~Riordan,
  \href{http://ebooks.cambridge.org/ref/id/CBO9781139167383}{{Percolation}},
  Cambridge University Press, Cambridge, 2006.
\newblock \href {https://doi.org/10.1017/CBO9781139167383}
  {\path{doi:10.1017/CBO9781139167383}}.

\bibitem{BY13}
B.~B{\l}aszczyszyn, D.~Yogeshwaran,
  \href{http://projecteuclid.org/euclid.ejp/1465064297}{{Clustering and
  percolation of point processes}}, Electron. J. Probab. 18~(72) (2013) 1.
\newblock \href {https://doi.org/10.1214/EJP.v18-2468}
  {\path{doi:10.1214/EJP.v18-2468}}.

\bibitem{BY14}
B.~B{\l}aszczyszyn, D.~Yogeshwaran,
  \href{https://www.cambridge.org/core/product/identifier/S000186780000690X/type/journal{\_}article}{{On
  comparison of clustering properties of point processes}}, Adv. Appl. Probab.
  46~(1) (2014) 1--20.
\newblock \href {https://doi.org/10.1239/aap/1396360100}
  {\path{doi:10.1239/aap/1396360100}}.

\bibitem{GKP16}
S.~Ghosh, M.~Krishnapur, Y.~Peres,
  \href{http://projecteuclid.org/euclid.aop/1474462100}{{Continuum percolation
  for Gaussian zeroes and Ginibre eigenvalues}}, Ann. Probab. 44~(5) (2016)
  3357--3384.
\newblock \href {https://doi.org/10.1214/15-AOP1051}
  {\path{doi:10.1214/15-AOP1051}}.

\bibitem{Gil76}
D.~T. Gillespie,
  \href{https://www.sciencedirect.com/science/article/pii/0021999176900413
  https://linkinghub.elsevier.com/retrieve/pii/0021999176900413}{{A general
  method for numerically simulating the stochastic time evolution of coupled
  chemical reactions}}, J. Comput. Phys. 22~(4) (1976) 403--434.
\newblock \href {https://doi.org/10.1016/0021-9991(76)90041-3}
  {\path{doi:10.1016/0021-9991(76)90041-3}}.

\bibitem{Gil77}
D.~T. Gillespie, \href{https://pubs.acs.org/doi/abs/10.1021/j100540a008}{{Exact
  stochastic simulation of coupled chemical reactions}}, J. Phys. Chem. 81~(25)
  (1977) 2340--2361.
\newblock \href {https://doi.org/10.1021/j100540a008}
  {\path{doi:10.1021/j100540a008}}.

\bibitem{EC20}
R.~Erban, S.~J. Chapman,
  \href{https://www.cambridge.org/core/product/identifier/9781108628389/type/book}{{Stochastic
  Modelling of Reaction–Diffusion Processes}}, Cambridge Texts in Applied
  Mathematics, Cambridge University Press, 2020.
\newblock \href {https://doi.org/10.1017/9781108628389}
  {\path{doi:10.1017/9781108628389}}.

\bibitem{SA92}
D.~Stauffer, A.~Aharony,
  \href{https://www.taylorfrancis.com/books/9781482272376}{{Introduction to
  Percolation Theory}}, Taylor {\&} Francis, London, 1992.
\newblock \href {https://doi.org/10.1201/9781315274386}
  {\path{doi:10.1201/9781315274386}}.

\bibitem{MM12}
S.~Mertens, C.~Moore, \href{https://link.aps.org/doi/10.1103/PhysRevE.86.061109
  http://arxiv.org/abs/1209.4936
  http://dx.doi.org/10.1103/PhysRevE.86.061109}{{Continuum percolation
  thresholds in two dimensions}}, Phys. Rev. E 86~(6) (2012) 061109.
\newblock \href {https://doi.org/10.1103/PhysRevE.86.061109}
  {\path{doi:10.1103/PhysRevE.86.061109}}.

\bibitem{GS81}
E.~T. Gawlinski, H.~E. Stanley,
  \href{https://doi.org/10.1088/0305-4470/14/8/007
  https://iopscience.iop.org/article/10.1088/0305-4470/14/8/007}{{Continuum
  percolation in two dimensions: Monte Carlo tests of scaling and universality
  for non-interacting discs}}, J. Phys. A. Math. Gen. 14~(8) (1981) L291--L299.
\newblock \href {https://doi.org/10.1088/0305-4470/14/8/007}
  {\path{doi:10.1088/0305-4470/14/8/007}}.

\bibitem{BBA83}
I.~Balberg, N.~Binenbaum, C.~H. Anderson,
  \href{https://link.aps.org/doi/10.1103/PhysRevLett.51.1605}{{Critical
  behavior of the two-dimensional sticks system}}, Phys. Rev. Lett. 51~(18)
  (1983) 1605--1608.
\newblock \href {https://doi.org/10.1103/PhysRevLett.51.1605}
  {\path{doi:10.1103/PhysRevLett.51.1605}}.

\bibitem{CG85}
J.~L. Cardy, P.~Grassberger, \href{https://doi.org/10.1088/0305-4470/18/6/001
  https://iopscience.iop.org/article/10.1088/0305-4470/18/6/001}{{Epidemic
  models and percolation}}, J. Phys. A. Math. Gen. 18~(6) (1985) L267--L271.
\newblock \href {https://doi.org/10.1088/0305-4470/18/6/001}
  {\path{doi:10.1088/0305-4470/18/6/001}}.

\bibitem{QT99}
J.~Quintanilla, S.~Torquato,
  \href{http://aip.scitation.org/doi/10.1063/1.479890}{{Percolation for a model
  of statistically inhomogeneous random media}}, J. Chem. Phys. 111~(13) (1999)
  5947--5954.
\newblock \href {https://doi.org/10.1063/1.479890}
  {\path{doi:10.1063/1.479890}}.

\bibitem{QTZ00}
J.~Quintanilla, S.~Torquato, R.~M. Ziff,
  \href{https://doi.org/10.1088/0305-4470/33/42/104
  https://iopscience.iop.org/article/10.1088/0305-4470/33/42/104}{{Efficient
  measurement of the percolation threshold for fully penetrable discs}}, J.
  Phys. A. Math. Gen. 33~(42) (2000) L399--L407.
\newblock \href {https://doi.org/10.1088/0305-4470/33/42/104}
  {\path{doi:10.1088/0305-4470/33/42/104}}.

\bibitem{NZ01}
M.~E.~J. Newman, R.~M. Ziff,
  \href{https://link.aps.org/doi/10.1103/PhysRevE.64.016706}{{Fast Monte Carlo
  algorithm for site or bond percolation}}, Phys. Rev. E 64~(1) (2001) 016706.
\newblock \href {https://doi.org/10.1103/PhysRevE.64.016706}
  {\path{doi:10.1103/PhysRevE.64.016706}}.

\bibitem{QZ07}
J.~A. Quintanilla, R.~M. Ziff,
  \href{https://link.aps.org/doi/10.1103/PhysRevE.76.051115}{{Asymmetry in the
  percolation thresholds of fully penetrable disks with two different radii}},
  Phys. Rev. E 76~(5) (2007) 051115.
\newblock \href {https://doi.org/10.1103/PhysRevE.76.051115}
  {\path{doi:10.1103/PhysRevE.76.051115}}.

\bibitem{BKL75}
A.~B. Bortz, M.~H. Kalos, J.~L. Lebowitz,
  \href{https://linkinghub.elsevier.com/retrieve/pii/0021999175900601}{{A new
  algorithm for Monte Carlo simulation of Ising spin systems}}, J. Comput.
  Phys. 17~(1) (1975) 10--18.
\newblock \href {https://doi.org/10.1016/0021-9991(75)90060-1}
  {\path{doi:10.1016/0021-9991(75)90060-1}}.

\bibitem{MK2}
{Machiko Katori}, {Makoto Katori},
  \href{http://arxiv.org/abs/2105.04142}{{Spreading and suppression of
  infection clusters on the Ginibre continuum percolation clusters}}, arXiv
  (2021).
\newblock \href {http://arxiv.org/abs/2105.04142} {\path{arXiv:2105.04142}}.

\bibitem{Hin00}
H.~Hinrichsen,
  \href{http://www.tandfonline.com/doi/abs/10.1080/00018730050198152}{{Non-equilibrium
  critical phenomena and phase transitions into absorbing states}}, Adv. Phys.
  49~(7) (2000) 815--958.
\newblock \href {https://doi.org/10.1080/00018730050198152}
  {\path{doi:10.1080/00018730050198152}}.

\bibitem{VFM09}
T.~Vojta, A.~Farquhar, J.~Mast,
  \href{https://link.aps.org/doi/10.1103/PhysRevE.79.011111}{{Infinite-randomness
  critical point in the two-dimensional disordered contact process}}, Phys.
  Rev. E 79~(1) (2009) 011111.
\newblock \href {https://doi.org/10.1103/PhysRevE.79.011111}
  {\path{doi:10.1103/PhysRevE.79.011111}}.

\bibitem{MS14}
N.~Miyoshi, T.~Shirai,
  \href{https://www.cambridge.org/core/product/identifier/S0001867800007394/type/journal{\_}article}{{A
  cellular network model with Ginibre configured base stations}}, Adv. Appl.
  Probab. 46~(3) (2014) 832--845.
\newblock \href {https://doi.org/10.1239/aap/1409319562}
  {\path{doi:10.1239/aap/1409319562}}.

\bibitem{LBDA15}
Y.~Li, F.~Baccelli, H.~S. Dhillon, J.~G. Andrews,
  \href{https://ieeexplore.ieee.org/document/7155510/}{{Statistical modeling
  and probabilistic analysis of cellular networks with determinantal point
  processes}}, IEEE Trans. Commun. 63~(9) (2015) 3405--3422.
\newblock \href {https://doi.org/10.1109/TCOMM.2015.2456016}
  {\path{doi:10.1109/TCOMM.2015.2456016}}.

\bibitem{MS16}
N.~Miyoshi, T.~Shirai,
  \href{https://www.jstage.jst.go.jp/article/transcom/E99.B/11/E99.B{\_}2016NEI0001/{\_}article}{{Spatial
  modeling and analysis of cellular networks using the Ginibre point process: A
  tutorial}}, IEICE Trans. Commun. E99.B~(11) (2016) 2247--2255.
\newblock \href {https://doi.org/10.1587/transcom.2016NEI0001}
  {\path{doi:10.1587/transcom.2016NEI0001}}.

\bibitem{TI06}
H.~Tamura, K.~R. Ito,
  \href{http://link.springer.com/10.1007/s00220-005-1507-2}{{A canonical
  ensemble approach to the fermion/boson random point processes and its
  applications}}, Commun. Math. Phys. 263~(2) (2006) 353--380.
\newblock \href {https://doi.org/10.1007/s00220-005-1507-2}
  {\path{doi:10.1007/s00220-005-1507-2}}.

\bibitem{TI07}
H.~Tamura, K.~R. Ito,
  \href{https://linkinghub.elsevier.com/retrieve/pii/S0022123606004411}{{A
  random point field related to Bose–Einstein condensation}}, J. Funct. Anal.
  243~(1) (2007) 207--231.
\newblock \href {https://doi.org/10.1016/j.jfa.2006.10.014}
  {\path{doi:10.1016/j.jfa.2006.10.014}}.

\bibitem{AFY19}
I.~Armend{\'{a}}riz, P.~A. Ferrari, S.~Yuhjtman,
  \href{http://arxiv.org/abs/1906.11120}{{Gaussian random permutation and the
  boson point process}}, arXiv (2019).
\newblock \href {http://arxiv.org/abs/1906.11120} {\path{arXiv:1906.11120}}.

\end{thebibliography}

\end{document}